\documentclass{iopart}

\usepackage{epsfig}
\usepackage{times}
\usepackage{color}

\newcommand{\Label}[1]{\label{#1}}
\newcommand{\rw}[1]{}
\def\DRAFT{\renewcommand{\Label}[1]{\label{##1}{\hbox to 0cm{\textcolor{green}{\hss\em ##1\quad}}}}
\renewcommand{\rw}[1]{\vskip 10pt\noindent{\framebox{\textcolor{red}{New Material Needed}}}\par\noindent{\textcolor{red}{\em ##1}}\vskip 10pt}
\def\Input##1{\include{##1}}}

\newcommand{\GPE}{Gross-Pitaevskii equation}
\newcommand{\BEC}{Bose-Einstein condensate}
\newcommand{\GP}{Gross-Pitaevskii}

\newcommand{\cg}[1]{{\textcolor{black}{#1}}}

\begin{document}

\title{The stochastic Gross-Pitaevskii equation.}
\author{C.W.~Gardiner $^{1}$ J.R. Anglin$^{2}$ and T. I. A. Fudge$^3$}

\begin{abstract}
We show how to adapt the ideas of local energy and momentum conservation in
order to derive modifications to the Gross-Pitaevskii equation which can be
used phenomenologically to describe irreversible effects in a Bose-Einstein
condensate. Our approach involves the derivation of a simplified 
quantum kinetic theory, in which all processes are treated locally.  
It is shown that this kinetic theory can then be transformed into a 
number of phase-space representations, of which the 
Wigner function description, although approximate, is shown to be the 
most advantageous.  In this description, the quantum kinetic master 
equation takes the form of a \GPE\ with noise and damping added 
according to a well-defined prescription---an equation we call the 
{\em stochastic \GPE}. From this, a very 
simplified description we call the {\em phenomenological growth equation}
can be derived.  We use this equation to study i) the nucleation and 
growth of vortex lattices, and ii) nonlinear losses in a 
hydrogen condensate, which it is shown can lead to a curious 
instability phenomenon. 

\end{abstract}

\address{$^{1}$School of Chemical and Physical Sciences, Victoria University, 
Wellington, New 
Zealand}

\address{$ ^2${Center for Ultracold
Atoms,
MIT 26-251, 77 Massachusetts Ave., Cambridge, MA 02139}
} 
\address{$ ^3${Wilton Research Institute,
Wilton, Wellington, New Zealand}
}

\section{Introduction}
\Label{secI}
There is a growing consensus that in some sense the dynamics of a 
trapped \BEC\ can be properly described by the \GPE, provided 
appropriate statistical assumptions are introduced.  This was first 
suggested by Svistunov 
\cite{Svistunov1991ax,Kagan1992axR,Kagan1992ax}, and numerical 
experiments were carried out by Damle and Sachdev \cite{Damle1996ax}, 
Marshall, New and Burnett \cite{Marshall1992ax}, and most recently by 
Davis \etal \cite{Davis2000bx}.  All of these considered the \GPE\ for 
a Bose gas inside a box, with random initial conditions, such that the 
total energy and number of particles would correspond to those of a 
partially condensed system in thermal equilibrium.  The numerical 
experiments then verified that this initial ensemble developed with 
time into a thermodynamic equilibrium ensemble as a result of the time 
evolution induced on each member of the ensemble by the \GPE. The 
recent work of Davis \etal \cite{Davis2000bx} also checked the 
dynamics of the resulting equilibrium state against Morgan's 
\cite{Morgan2000ax} theoretical predictions from quantum field theory, 
and found that these were in agreement with each other.

The work of Steel {\em et al.} \cite{Steel1998b} took a different, and more 
rigorous view, that the correct physical relationship between the condensate 
wavefunction with random initial conditions and the physical field operator is 
made by means of a truncated Wigner function representation.  It is known 
\cite{Gardiner1999ax} that for the standard \BEC\ Hamiltonian, the Wigner 
function obeys a 
generalized Fokker-Planck equation, in which derivatives up to third order 
occur. However, if the amplitudes of the Wigner function phase-space variables 
are large, the third order derivative terms are small, and can be 
neglected---this ``truncation'' gives the method its name.  However, we also 
find that the expected second-order derivative terms, which correspond to 
noise, are for this Hamiltonian exactly zero, and thus what remains corresponds 
to a Liouville equation for an ensemble of wavefunctions which obey the \GPE. 

However the description can only be valid provided the amplitudes can be 
regarded as large, and this must be the case for all modes.  Clearly, the 
higher energy (or smaller wavelength) modes will always have a very small 
amplitude, and cannot be described by a truncated Wigner function. Indeed, if 
no truncation is done, then the result of the quantum mechanical 
constraints---essentially Heisenberg's uncertainty principle---on the Wigner 
function leads to a delta function spatial correlation function, corresponding 
to infinite fluctuations at each point in space, which cannot be simulated.  In 
all simulations, however, the implementation of the spatial differentiation
requires a spatial grid, whose spacing corresponds to the inverse of the 
highest spatial frequency used, thus providing a cutoff in both space and 
energy.  The results must then be cutoff dependent.

The work of Davis \etal\cite{Davis2000bx} is the first to put down 
explicitly that a cutoff in energy or spatial frequency must be 
imposed on this \GPE\ equation.  However, the basic idea, that high 
and low energies require different treatments is that which forms the 
basis of our quantum kinetic theory, which gives a formalism for 
coupling together a fully thermalized noncondensate band, consisting 
of particle with energies more than a certain value $ E_R$, to a 
condensate band, consisting of all lower energy particles.  The 
non-condensate band can be treated by the quantum Boltzmann equation, 
while the condensate band should in principle be treated fully quantum 
mechanically.

In this paper we will combine the Wigner function ideas of Steel {\em 
et al.} \cite{Steel1998b} with quantum kinetic theory, and as well 
introduce the idea of local energy conservation as suggested by 
Zaremba {\em et al.} \cite{Zaremba1998a,Zaremba1999a} to 
derive a method of coupling the thermalized noncondensate band to the 
\GPE. This results in a \GPE\ with added noise---a stochastic \GPE.

The paper consists of three main parts: Sect.\ref{secII} implements 
the idea of local energy and momentum conservation in quantum kinetic 
theory and results in a relatively simple master equation for the 
interaction of a condensate with a fixed bath of non-condensed atoms; 
Sect.\ref{secIII} shows how to implement the Wigner function method to 
transform the master equation into a c-number equation with random 
initial conditions, and to compare with P- and Q-function methods; 
Sect.\ref{secIV} develops a very much simplified phenomenological 
equation, which we show can be applied to the two-body loss problem in 
the hydrogen condensate experiments, and to the stabilization of
quantized vortex arrays.  In Sect.\ref{secV} we adapt the 
phenomenological growth equation to include losses from dipolar 
relaxation in a hydrogen condensate, and show that the nonlinear 
losses so introduced could lead to a ``boom and bust'' instability in 
which the condensate grows and collapses repeatedly, although the 
conditions under which hydrogen condensates are presently formed make 
it difficult to say whether this could be observed in practice.

\section{Application of local energy conservation to quantum kinetic theory}
\Label{secII}
In a formulation of the kinetic theory of non-condensed vapour in 
interaction with a condensate, Zaremba, Nikuni and Griffin 
\cite{Zaremba1999a} have used the concepts of local energy and 
momentum 
conservation to develop appropriate equations of motion for the system 
in a formulation based on Hartree-Fock-Popov methodology 
\cite{Griffin1996b,Fetter1972ax,Proukakis1998ax,Dalfovo1999a}.  
This is in contrast to our own 
\cite{Gardiner1997a,Anglin1997a,Gardiner1997b,Jaksch1997ax,Gardiner1998a,%
Gardiner1998b,Jaksch1998a,Gardiner2000a} formulation of quantum 
kinetic theory, in which energy conservation is expressed in terms of 
transitions between eigenstates of the condensate.  While it is clear 
that a description in terms of eigenfunctions must be more accurate, 
it yields equations which are not easy to handle exactly, and thus the 
greater accuracy can in practice be an illusion.

The full formulation of our quantum kinetic theory is to be found in QKV
\footnote{\noindent In this paper will use the notation:
QKI   for \cite{Gardiner1997a},
QKII  for \cite{Jaksch1997ax},
QKIII for \cite{Gardiner1998a}, 
QKIV  for \cite{Jaksch1998a},
QKV   for \cite{Gardiner2000a},
QKVI   for \cite{Lee2000ax} and
QKVII   for \cite{Davis2000ax}}, which 
depends strongly on  QKIII.  The major improvement in QKV is a full 
consideration of the mean field effects of condensate on vapour and conversely, 
but the general methodology is very similar.  In these we divide the excitation 
spectrum at an energy $ E_R$, above which the excitations can be treated as 
particle like. The field operator is written (in the Schro\"dinger picture) as
\begin{eqnarray}\Label{qk1}
\psi({\bf x}) = \psi_{\rm NC}({\bf x}) + \phi({\bf x})
\end{eqnarray}
where the first operator represents the modes with excitation energies above 
$ E_R$ (and is called the noncondensate band field operator), while 
$ \phi({\bf x})$ represents the excitations with energies below $ E_R$
(and is called the condensate band field operator).  This division is chosen 
and fixed for any given problem, and is to some extent arbitrary, since it is 
expected that the higher modes of the condensate band will be very close to 
particle-like, and in practice will also be thermalized.

Let us look at the derivation of energy conservation used in QKIII, using the 
notation used there. On p539
we  examine a term in the master equation involving $ H^{(1)}_{I,C}$ as 
defined in (QKIII.8), which contains one
$ \phi({\bf x})$ or $ \phi^\dagger({\bf x})$, where $ \phi({\bf x})$
is the field operator for the condensate band.
Explicitly, the term in Hamiltonian can be written
\begin{eqnarray}\Label{bba701}
H^{(1)}_{I,C}&=&
\int d^3{\bf x}\,Z_{3}({\bf x})\phi^\dagger({\bf x})
+ {\rm h.c.}
\end{eqnarray}
In this equation we have defined a notation
\begin{eqnarray}\Label{bba802}
Z_{3}({\bf x}) 
&=& u \psi^\dagger_{{\rm NC}} ({\bf x})\psi_{{\rm NC}} ({\bf x})\psi_{{\rm NC}} 
({\bf x}).
\end{eqnarray}
Substituting into the master equation (QKIII.20), terms arise which are of the 
form
\begin{eqnarray}\Label{bba8}
&&-{1\over\hbar^2}\int d^3{\bf x}\int d^3{\bf x'}\int_0^\infty d\tau\,
{\rm Tr}\left\{Z_{3}({\bf x})Z^\dagger_{3}({\bf x}',-\tau)\rho_{{\rm NC}}\right
\}
\nonumber \\
&&\qquad \times \phi^\dagger({\bf x}) \phi({\bf x'},-\tau)
\rho_{\rm C}(t).
\end{eqnarray}
Here we have used the notation
\begin{eqnarray}\Label{notations1}
Z_3({\bf x},t)& =& e^{iH_{{\rm NC}}t/\hbar}Z_3({\bf x}) e^{-iH_{{\rm NC}}t/
\hbar}
\\ \Label{notations2}
\phi({\bf x},t)& =& e^{iH_{0}t/\hbar}\phi({\bf x}) e^{-iH_{0}t/\hbar}.
\end{eqnarray}
In QKIII we expanded the operators $ \phi({\bf x}) $ in eigenoperators of the 
condensate band Hamiltonian $ H_{0} $, so as to be able to perform the integral 
over the time $ \tau$, and thus arrive at the final master equation, which 
would not involve $ \phi$ operators at different times.  This method does not 
require that there be a condensate.  If we assume that there is a condensate 
present, we use this knowledge to make an estimate of the time development in 
the time $ \tau$.

\subsection{Local energy conservation}
The basis for this concept is given by the works of Zaremba, Nikuni, Griffin, 
and co-workers 
\cite{Zaremba1998a,Zaremba1999a,Williams2001b}, which 
themselves draw on the work of Kirkpatrick and Dorfman 
\cite{Kirkpatrick1983a,Kirkpatrick1985b,Kirkpatrick1985c}, and can be 
seen phenomenologically by use of the density-phase description of the 
Gross-Pitaevskii equation
\begin{eqnarray}\Label{1}
i\hbar{\partial\xi({\bf x},t) \over\partial t}
&=& -{\hbar^2\over 2m}\nabla^2\xi({\bf x},t) 
+V_T({\bf x})\xi({\bf x},t) + u \bigl |\xi({\bf x},t)\bigr |^2 
\xi({\bf x},t) .
\end{eqnarray}
By writing now 
\begin{eqnarray}\Label{2}
\xi({\bf x},t) &=& \sqrt{n_{\rm C}({\bf x},t)}\exp[i\Theta({\bf x},t)]
\end{eqnarray}
the \GP\ equation takes the form
\begin{eqnarray}\Label{3a}
{\partial {n_{\rm C}({\bf x},t)} \over\partial t } &=&
- \nabla\cdot[{{\bf v}_{\rm C}({\bf x},t)}{n_{\rm C}({\bf x},t)}],
\\ \Label{3b}
\hbar{\partial\Theta({\bf x},t) \over\partial t} 
&=& -\mu_{\rm C}({\bf x},t) -{1\over 2}mv_{\rm C}({\bf x},t)^2 
\\ \Label{3c}
&\equiv & -\epsilon_{\rm C}({\bf x},t).
\end{eqnarray}
in which the {\em condensate velocity} $ {\bf v}_{\rm C}({\bf x},t)$, and the 
{\em local chemical potential} $ \mu_{\rm C}({\bf x},t) $ are given by
\begin{eqnarray}\Label{4a}
{\bf v}_{\rm C}({\bf x},t) &\equiv& {\hbar\over m}\nabla\Theta({\bf x},t),
\\ \Label{4b}
\mu_{\rm C}({\bf x},t)&\equiv&
-{\hbar^2\nabla^2\sqrt{n_{\rm C}({\bf x},t)}\over \sqrt{n_{\rm C}({\bf x},t)}} 
+V_T({\bf x}) + u\,n_{\rm C}({\bf x},t).
\end{eqnarray}
This transcription of the \GP\ equation is the appropriate basis for a 
description of condensate behaviour when the condensate density 
$n_{\rm C}({\bf x},t)$ is slowly varying in space and time, so that we may 
consider 
the dependence on space and time to arise largely from the phase 
$ \Theta({\bf x},t)$.  This leads to the {\em hydrodynamic approximation} in 
which the Laplacian term in (\ref{4b}) is dropped and thence to the energy 
conservation equation
\begin{eqnarray}\Label{401}
m{\partial{\bf v}_{\rm C}({\bf x},t) \over\partial t} &=& -\nabla\left[V_T({\bf 
x})
+{1\over 2}mv_{\rm C}({\bf x},t)^2 
+un_{\rm C}({\bf x},t)\right]
\end{eqnarray}
The static solution of the  
hydrodynamic description is the Thomas-Fermi description of the condensate 
wavefunction, in which 
\begin{eqnarray}\Label{5a}
{\bf v}_{\rm C}({\bf x},t) &=& 0,
\\ \Label{5b}
\mu_{\rm C}({\bf x},t) &=& \mu,
\\ \Label{5c}
n_{\rm C}({\bf x},t) &=& {\mu - V_T({\bf x})\over u},
\\ \Label{5d}
\Theta({\bf x},t) &=& -{\mu t\over\hbar}.
\end{eqnarray}
In the hydrodynamic regime, the local chemical potential $ \mu_{\rm C}({\bf 
x},t)$ is 
a strictly local quantity, and is a simple function of the density 
$ n_{\rm C}({\bf  x},t)$.  The local energy density $ \epsilon_{\rm C}({\bf 
x},t)$, 
defined 
in (\ref{3c}) is explicitly equal to the derivative of the energy 
density (also evaluated using the hydrodynamic approximation)
\begin{eqnarray}\Label{6a}
E({\bf x},t) &=& {1\over2}mv_{\rm C}({\bf x},t)^2 n_{\rm C}({\bf x},t)
+V_T({\bf x})n_{\rm C}({\bf x},t) +{1\over 2} n_{\rm C}({\bf x},t)^2
\end{eqnarray}
with respect to $  n_{\rm C}({\bf x},t) $.  Thus, when considering the possible 
addition of a particle to the condensate, it is intuitively appealing to 
say that this takes place at a position $ {\bf x}$, and 
that the energy added to the condensate by this process is 
$\epsilon({\bf x},t) $.

This is given some further backing by the knowledge that energy conservation in 
quantum mechanics depends on frequency matching, and that to the extent that 
$ n_{\rm C}({\bf x},t)$ and $ \epsilon({\bf x},t) $ depend slowly on time, the 
relevant frequency will be $\epsilon({\bf x},t). $
If $ n_{\rm C}({\bf x},t)$ and $ {\bf v}_{\rm C}({\bf x},t)$ are both slowly 
dependent on 
space, the wavelength of the wavefunction will correspond to a momentum
$ m{\bf v}_{\rm C}({\bf x},t)$, and thus momentum conservation will involve 
this 
momentum.

\subsection{Application to quantum kinetic theory}
The major task in developing a quantum stochastic description is to 
find a simple way of expressing the term $ \phi({\bf x}',-\tau)$ in 
terms of $ \phi({\bf x},0)$, and we would like to follow Zaremba \etal 
\cite{Zaremba1999a} in doing this.  We assume that the evolution over 
this short time can be approximated by simply the phase evolution of 
the condensate according to (\ref{3b},\ref{3c}).  We also want to 
express the result in terms of operators at the same location, which 
we take as the midpoint $ {\bf u} =({\bf x}+{\bf x'})/2$.  Thus, using 
the symbol $ {\bf y}={\bf x}- {\bf x}'$, we write
\begin{eqnarray}\fl
\Label{10}
 \phi^\dagger({\bf x}) \phi({\bf x'},-\tau) 
&\approx& \phi^\dagger({\bf u}) \phi({\bf u})
\exp\{i [-\Theta({\bf y}/2,t) + \Theta(-{\bf y}/2,t-\tau)]\}
\\ \Label{11}
&\approx & \phi^\dagger({\bf u}) \phi({\bf u})
\exp\left\{{i\over\hbar}[
-m{\bf v}_{\rm C}({\bf u},t)\cdot{\bf y} -\epsilon_{\rm C}({\bf u},t)\tau]
\right
\}.
\end{eqnarray}
There are two main assumption here:
\begin{itemize}
\item[i)] There is an assumption that the state of the condensate band 
is dominated by a single condensate wavefunction.  However, the 
wavefunction may have any form, and of course may also be time 
dependent.  It is possible that a more accurate method could be 
developed using hydrodynamic quantization, such as that developed by 
Marques and Bagnato \cite{Marques2000a}, which yields operator 
equations almost the same as the standard hydrodynamic formalism 
outlined above.  
\item[ii)] The approximation (\ref{11}) requires that 
only small $ {\bf y}$ and $ \tau$ contribute.  This will be valid when 
the function $ {\rm Tr}\left\{Z_3({\bf x})Z^\dagger_3({\bf 
x}',-\tau)\rho_{\rm NC}\right\}$, which is convoluted with $ 
\phi^\dagger({\bf x}) \phi({\bf x'},-\tau)$ in (\ref{bba8}), is almost 
local in space and $ \tau$.  This kind of behaviour is expected if the 
noncondensate band is fully thermalized at a sufficiently high 
temperature.
\end{itemize}
Both of these assumptions are used by Zaremba \etal \cite{Zaremba1999a}

\subsection{\Label{Sect. 6.5.2}Final form of the master equation}
By carrying out the procedures used in QKIII, we can 
finally get the master equation:
\begin{eqnarray}
\fl\Label{12a}
\dot \rho _{\rm C}( t)  ={i\over\hbar}   \int d^3{\bf x}\,\Bigg[
\phi^{\dagger}( {\bf x})\left({\hbar^2\nabla^2\over2m}-V_T({\bf x})
-2u \rho_{\rm NC}({\bf x},t)
 -{1\over 2}u\phi^\dagger({\bf x})\phi({\bf x})\right)
\phi( {\bf x})\, ,\,\rho_{\rm C} \Bigg] 
\nonumber\\
\\ \fl
 \Label{12b}
+\int d^3{\bf x}\Bigg(
G^{(+)}[{\bf x},\epsilon_{\rm C}({\bf x},t)]
\left(2\phi ({\bf x})\rho_{\rm C}\phi^\dagger ({\bf x})
-\rho_{\rm C}\phi^\dagger ({\bf x})\phi ({\bf x})
-\phi^\dagger ({\bf x})\phi ({\bf x})\rho_{\rm C}\right)
\\ \fl \Label{12c}
\qquad\quad\,  +
G^{(-)}[{\bf x},\epsilon_{\rm C}({\bf x},t)]
\left(2\phi^\dagger ({\bf x})\rho_{\rm C}\phi ({\bf x})
-\rho_{\rm C}\phi ({\bf x})\phi^\dagger ({\bf x})
-\phi ({\bf x})\phi^\dagger ({\bf x})\rho_{\rm C}\right)
\\ \fl \Label{12d} 
\qquad\quad\, 
+ M[{\bf x}]
\left(2U({\bf x})\rho_{\rm C}U^\dagger({\bf x})
-\rho_{\rm C}U^\dagger({\bf x})U({\bf x})
-U^\dagger({\bf x})U({\bf x})\rho_{\rm C}\right)
\\  \fl	\Label{12e}
\qquad\quad\, 
+ E^{(+)}[{\bf x},\epsilon_{\rm C}({\bf x},t)]
\left(2V({\bf x})\rho_{\rm C}V^\dagger({\bf x})
-\rho_{\rm C}V^\dagger({\bf x})V({\bf x})
-V^\dagger({\bf x})V({\bf x})\rho_{\rm C}\right)
\\ \fl \Label{12f}
\qquad\quad\, 
+
E^{(-)}[{\bf x},\epsilon_{\rm C}({\bf x},t)]
\left(2V^\dagger({\bf x})\rho_{\rm C}V({\bf x})
-\rho_{\rm C}V({\bf x})V^\dagger({\bf x})
-V({\bf x})V^\dagger({\bf x})\rho_{\rm C}\right)\Bigg).
\end{eqnarray}
For compactness, we have defined
\begin{eqnarray}\Label{20}
U({\bf x}) &\equiv& \phi^\dagger({\bf x})\phi({\bf x}),
\\ \Label{21}
V({\bf x}) &\equiv& \phi^\dagger({\bf x})\phi({\bf x})\phi({\bf x}),
\\ \Label{22}
\rho_{\rm NC}({\bf x},t) &\equiv&\int d^3{\bf K}\,F({\bf K},{\bf x}).
\end{eqnarray}
Here $F({\bf K},{\bf x})$ is the phase space density of the 
noncondensate band in terms of wavenumber ${\bf K}$ and position 
${\bf x}$.

The transition rates 
$ E^{(\pm ) }$, $ M $, $ G^{(\pm ) } $ are defined by
\begin{eqnarray}\fl\Label{23a}
G^{( +) }[ {\bf x},\epsilon] =
\frac{ u^2}{(2\pi)^5\hbar ^2}
\int d^3{\bf K}_1\int d^3{\bf K}_2\int d^3{\bf K}_3
F({\bf K}_1,{\bf x})
F({\bf K}_2,{\bf x})[F({\bf K}_3,{\bf x})+1]
\nonumber \\ \fl
\qquad\qquad\qquad\qquad\qquad\qquad\times
\delta\left(\Delta{\bf K}_{123}-{m{\bf v}_{\rm C}({\bf x},t)\over\hbar}\right)
 \delta\left(\Delta\omega_{123}-{\epsilon\over\hbar}\right)  
 \\
\fl
  \Label{23b}
G^{( -) }[ {\bf x},\epsilon] = 
 \frac{ u^2}{(2\pi)^5\hbar ^2}
\int d^3{\bf K}_1\int d^3{\bf K}_2\int d^3{\bf K}_3
[F({\bf K}_1,{\bf x})+1]
[F({\bf K}_2,{\bf x})+1]F({\bf K}_3,{\bf x})
\nonumber \\
\fl
\qquad\qquad\qquad\qquad\qquad\qquad \times
 \delta\left(\Delta{\bf K}_{123}-{m{\bf v}_{\rm C}({\bf x},t)\over\hbar}\right)
 \delta\left(\Delta\omega_{123}-{\epsilon\over\hbar}\right)  
\\ \fl
\Label{23c}
M({\bf x}) = 
\frac{2 u^2}{(2\pi)^2\hbar ^2}
\int d^3{\bf K}_1\int d^3{\bf K}_2
F({\bf K}_1,{\bf x})
[F({\bf K}_2,{\bf x})+1]
\delta({\bf K}_1-{\bf K}_2)
 \delta(\omega_1 -\omega _2)  
\\ \fl
\Label{23d}
E^{( +) }({\bf x},\epsilon) =
2\pi\frac{ u^2}{2\hbar ^2}
\int d^3{\bf K}_1 
F({\bf K}_1,{\bf x})
 \delta\left(\omega_1 -{\epsilon\over\hbar}\right)  
\delta\left({{\bf K}_1-{m{\bf v}_{\rm C}({\bf x},t)\over\hbar}}\right)
 \\ \fl
 \Label{23e}
E^{( -) }({\bf x},\epsilon) = 
 2\pi\frac{ u^2}{2\hbar ^2}
\int d^3{\bf K}_1
[F({\bf K}_1,{\bf x})+1]
 \delta\left(\omega_1 -{\epsilon\over\hbar}\right)  
\delta\left({{\bf K}_1-{m{\bf v}_{\rm C}({\bf x},t)\over\hbar}}\right)
\end{eqnarray}
In the above equations we use the notations
\begin{eqnarray}
\Label{omegaku}
\hbar\omega({\bf K},{\bf x})&\equiv& {\hbar^2{\bf K}^2\over 2m}
+V_T({\bf 
x})
\\
\Label{omega1}
\omega_i &\equiv& \omega({\bf K}_i,{\bf x})
\\ \Label{deltaK}
\Delta{\bf K}_{123}&\equiv& {\bf K}_{1}+{\bf K}_{2}-{\bf K}_{3}
\\
\Label{deltaOmega}
\Delta\omega_{123} &\equiv& \omega_{1}+\omega_{2}-\omega_{3}.
\end{eqnarray}

\subsubsection{Effect of the cutoff at $ E_R$}
The fact that $ \phi({\bf x})$ has an upper energy cutoff means that the 
commutator $ [\phi({\bf x}),\phi^\dagger({\bf x}')]$ is only an approximate 
delta function.  This has no influence on the form of the master equation, but 
equations of motion derived from it for averages of $ \phi$ will automatically
ensure that solutions remain in the correct subspace.
\subsubsection{Conservation of energy and momentum}\Label{Conservation}

The choice made in Eqs (\ref{10},\ref{11}) is very drastic, assigning as it 
does a single phase to all condensate band modes, and this has the consequence 
that the terms (\ref{12d}--\ref{12f}) are in fact zero.  For the last two this 
is obvious, since the energy and momentum conservation delta functions in the 
definitions of $ E^{(\pm)}$ would require the equality of the energy and 
momentum of a particle in the condensate band with those of a particle in the 
noncondensate band, which is by definition not possible.
The vanishing of the term (\ref{12d}) is not so immediately obvious, but 
momentum and energy conservation here mean that in 
fact this is a forward scattering term, which the methodology of QKIII and QKV 
explicitly removes from the irreversible part.

The origin of this is the requirement that, at a given position $ {\bf x}$,
there is a unique energy and momentum which a particle in the condensate band 
can have.  This is actually only true for particles contained in the 
condensate 
itself, so that we can expect that there is a spread of values of 
$ \epsilon_{\rm C}({\bf x},t)$ and $ {\bf v}_{\rm C}({\bf x},t)$.  As well as 
this, it 
should be borne in mind that the hydrodynamic approximation is the foundation 
of these results, and this is indeed an approximation.  Inclusion of 
corrections to this approximation would also yield a spread in local momenta 
and energies available for particles in the condensate band.

The effect of including a spread in values of $ \epsilon_{\rm C}({\bf x},t)$ 
and 
$ {\bf v}_{\rm C}({\bf x},t)$ would be principally to give a nonzero value to 
the 
terms (\ref{12d}--\ref{12f}) in the master equation.  The last two terms would 
certainly be very small, but the scattering term (\ref{12d}) should become 
significant.  The growth terms (\ref{12a},\ref{12b}) would be modified, but not 
greatly.  However the changes are only in the nonoperator coefficients---the 
basic structure of the master equation remains the same.
The major point of simplification compared with the treatment of QKIII and QKV 
is the expression of the irreversible terms directly in terms of the field 
operators $ \phi({\bf x})$, which enables a much simpler analysis to be 
carried 
out.
\subsubsection{Time development equations for the noncondensate band}
The noncondensate band is described by the phase space density 
$ F({\bf K}, {\bf x})$, whose equation of motion is given in QKV, in the form 
of
a quantum Boltzmann equations and with added terms to take account of transfer 
between the condensate and noncondensate bands.
For simplicity, in this paper we assume that the noncondensate band is fully 
thermalized with temperature $ T$ and chemical potential $ \mu$.

\section{Equivalent Wigner function treatment}
\Label{secIII}
\subsection{Stochastic differential equations for the Wigner function}
A master equation of the form we have produced can be treated using phase 
space representation methods \cite{Gardiner1999ax}. These will give rise to 
equivalent 
stochastic differential equations for a c-number phase space variable 
$ \alpha({\bf x},t)$ whose averages are related to those of the operator 
$ \phi({\bf x})$ using symmetric ordering: thus
\begin{eqnarray}\Label{28a}
\langle \alpha({\bf x},t)\rangle &=& \langle \phi({\bf x})\rangle,
\\ \Label{28b}
\langle \alpha^*({\bf x},t)\rangle &=& \langle \phi^\dagger({\bf x})\rangle,
\\ \Label{28c}
\langle \alpha^*({\bf x},t) \alpha({\bf x}',t)\rangle
 &=& {\langle \phi^\dagger({\bf x})\phi({\bf x})
+\phi({\bf x}')\phi^\dagger({\bf x})\rangle\over 2}.
\end{eqnarray}
Using the operator correspondences in \cite{Gardiner1999ax} Eq. (4.5.12), we 
can deduce the 
stochastic differential equation
\begin{eqnarray}\fl\Label{29}
{d\alpha({\bf x},t) } =
{i\over\hbar}  \left\{
{\hbar^2\over2m}\nabla^2\alpha( {\bf x},t)-V_T({\bf x})\alpha( {\bf x},t)
-2u \rho_{\rm NC}({\bf x},t)\alpha( {\bf x},t)
 -u|\alpha({\bf x})|^2\alpha( {\bf x})
\right\}\,dt 
\nonumber\\ \fl \qquad\qquad
+ \left\{ G^{(+)}[{\bf x},\epsilon_{\rm C}({\bf x},t)]
- G^{(-)}[{\bf x},\epsilon_{\rm C}({\bf x},t)]
          -  M[{\bf x}]\right\} \alpha({\bf x},t)\,dt
\nonumber \\ \fl\qquad\qquad
+  dW_{G}({\bf x},t) +  i\alpha({\bf x})dW_{M}({\bf x},t) .
\end{eqnarray}
Here the last two terms are Gaussian white noise terms.  The quantity 
$ dW_M({\bf x},t) $  is a real Wiener noise term, to be interpreted in the Ito 
sense \cite{Gardiner1989ax}, whose correlation functions 
are
\begin{eqnarray}\Label{30}
&& \langle dW_{M}({\bf x},t)\rangle=0,
\\ &&
\langle dW_{M}({\bf x},t) dW_{M}({\bf x}',t)\rangle
=2M[{\bf x}]\delta({\bf x}- {\bf x}')\,dt,
\end{eqnarray}
while the other noise is complex, with correlation functions
\begin{eqnarray}\fl
\Label{31}\langle dW_{G}({\bf x},t)\rangle=
\langle dW^*_{G}({\bf x},t)\rangle=0,
\\ \fl
\Label{W32}
\langle dW^*_{G}({\bf x},t) dW^*_{G}({\bf x}',t)\rangle=0,
\\ \fl
\Label{W33}
\langle dW_{G}({\bf x},t) dW_{G}({\bf x}',t)\rangle=0,
\\ \fl
\Label{W34} 
\langle dW^*_{G}({\bf x},t) dW_{G}({\bf x}',t)\rangle
={\delta({\bf x}- {\bf x}')\over 2}
\left( G^{(+)}[{\bf x},\epsilon_{\rm C}({\bf x},t)]
+ G^{(-)}[{\bf x},\epsilon_{\rm C}({\bf x},t)]\right) dt.
\end{eqnarray}
\subsubsection{Neglect of third order noise terms}
The equation is also approximate in that third order noise terms do arise along 
with the trilinear term.  These terms arise as third order partial derivatives 
in the Fokker-Planck equation generated by the Wigner function representation 
of the master equation (\ref{12a}--\ref{12f}), and have no straightforward 
stochastic interpretation.  When the variable 
$ \alpha({\bf x})$ is large, corresponding to large phase space occupation, 
these terms become small, as pointed out in \cite{Steel1998b}.

 
 \subsubsection{Stoof's equation}
 
 Stoof \cite{Stoof1999x} has developed a similar form of Wigner function 
 stochastic \GPE\ using path integral methods, and this has been used to 
 analyse damping of condensate modes and the reversible formation of a 
 condensate \cite{Stoof2001a,Duine2001a}.  The actual form of the damping 
 is not the same as ours, but it is easy to see that to lowest order in the 
 damping it is equivalent, and this is all we would claim for our 
 derivation.  The noise terms are equivalent to ours.
  
 In \cite{Stoof2001a} a case is considered where one can take a 
 ``classical'' limit of this Wigner approach---a low energy approximation 
 and essentially always valid for the description of the dynamics of the 
 condensate and its low-lying excitations---and in this approximation the 
 P- Q- and Wigner function forms of the noise become the same.  This leads 
 to an approximate noise term which is of the the same {\em form} as that 
 of the Q-function.  
 

\subsection{Stochastic differential equations for P- and Q-functions}
\subsubsection{Use of the P-function}
It is more conventional in quantum optics to use either the Glauber P-function 
or the positive P-function, instead of the Wigner function. The main advantage 
is that the approximation, necessary for the Wigner function, that third order 
noise terms must be neglected is not required---the resulting stochastic 
differential equations are exact.

For either 
P-function form, we would replace the symmetrized product  averaging rule with
one involving normal products, so that we would obtain
\begin{eqnarray}\Label{P28a}
\langle \alpha({\bf x},t)\rangle &=& \langle \phi({\bf x})\rangle,
\\ \Label{P28b}
\langle \alpha^*({\bf x},t)\rangle &=& \langle \phi^\dagger({\bf x})\rangle,
\\ \Label{P28c}
\langle \alpha^*({\bf x},t) \alpha({\bf x}',t)\rangle
 &=& {\langle \phi^\dagger({\bf x})\phi({\bf x})\rangle}.
\end{eqnarray}
Using the operator correspondences in \cite{Gardiner1989ax} Eq. (4.5.9), we can 
deduce the 
same stochastic differential equation (\ref{29}),
but with the correlation functions
\begin{eqnarray}\Label{P30}
&& \langle dW_{M}({\bf x},t)\rangle=0,
\\ &&
\langle dW_{M}({\bf x},t) dW_{M}({\bf x}',t)\rangle
=2M[{\bf x}]\delta({\bf x}- {\bf x}')\,dt,
\end{eqnarray}
(The same as for the Wigner function), while the other noise has the altered 
correlation functions
\begin{eqnarray}\Label{P31}
&&\langle dW_{G}({\bf x},t)\rangle=
\langle dW^*_{G}({\bf x},t)\rangle=0,
\\
\Label{P32}
 &&
\langle dW^*_{G}({\bf x},t) dW^*_{G}({\bf x}',t)\rangle
={iu\over\hbar}\alpha^*({\bf x})^2,
\\ \Label{P33}
&&
\langle dW_{G}({\bf x},t) dW_{G}({\bf x}',t)\rangle=-{iu\over\hbar}
\alpha^*({\bf x})^2,
\\ \Label{P34}&&
\langle dW^*_{G}({\bf x},t) dW_{G}({\bf x}',t)\rangle
={\delta({\bf x}- {\bf x}')}
 G^{(-)}[{\bf x},\epsilon_{\rm C}({\bf x},t)] dt.
\end{eqnarray}
The interpretation of (\ref{29}) as  a genuine stochastic differential equation  
requires that the matrix of noise coefficients
\begin{eqnarray}\Label{P36}
\left(\begin{array}{cc} G^{(-)}[{\bf x},\epsilon_{\rm C}({\bf x},t)]
 & -{iu\over\hbar}\alpha^*({\bf x})^2 
\\
{iu\over\hbar}\alpha({\bf x})^2 & 
 G^{(-)}[{\bf x},\epsilon_{\rm C}({\bf x},t)]
 \end{array}\right)
\end{eqnarray}
should have only nonnegative eigenvalues.  For higher temperature 
situations in which there is a substantial thermal component, this 
will certainly be true for all values of the variable $ \alpha({\bf 
x}) $ which would turn up in a stochastic simulation.  When this is 
not so, a Positive P-representation would be necessary.  The 
experience of Drummond and co-workers 
\cite{Drummond1999a,Drummond2000a} has shown this is in principle 
feasible, but application to experimentally realistic problems would 
be very difficult.

\subsubsection{Use of the Q-function}
In the case of the Q-function  
one should replace the symmetrized product  averaging rule with
one involving antinormal products, so that we would obtain
\begin{eqnarray}\Label{Q28a}
\langle \alpha({\bf x},t)\rangle &=& \langle \phi({\bf x})\rangle,
\\ \Label{Q28b}
\langle \alpha^*({\bf x},t)\rangle &=& \langle \phi^\dagger({\bf x})\rangle,
\\ \Label{Q28c}
\langle \alpha^*({\bf x},t) \alpha({\bf x}',t)\rangle
 &=& {\langle \phi({\bf x})\phi^\dagger({\bf x}')\rangle}.
\end{eqnarray}
Using the operator correspondences in \cite{Gardiner1999ax} Eq. (4.5.10), we 
can deduce 
again the 
same stochastic differential equation (\ref{29}),
but with the correlation functions
\begin{eqnarray}\Label{Q30}
&& \langle dW_{M}({\bf x},t)\rangle=0,
\\ &&
\langle dW_{M}({\bf x},t) dW_{M}({\bf x}',t)\rangle
=2M[{\bf x}]\delta({\bf x}- {\bf x}')\,dt,
\end{eqnarray}
(Again the same as for the Wigner function), while the other noise has the 
altered 
correlation functions
\begin{eqnarray}\Label{Q31}
&&\langle dW_{G}({\bf x},t)\rangle=
\langle dW^*_{G}({\bf x},t)\rangle=0,
\\ &&
\langle dW^*_{G}({\bf x},t) dW^*_{G}({\bf x}',t)\rangle
=-{iu\over\hbar}\alpha^*({\bf x})^2,
\\ \Label{Q32}
&&
\langle dW_{G}({\bf x},t) dW_{G}({\bf x}',t)\rangle= {iu\over\hbar}
\alpha^*({\bf x})^2,
\\ 
\Label{Q34}&&
\langle dW^*_{G}({\bf x},t) dW_{G}({\bf x}',t)\rangle
={\delta({\bf x}- {\bf x}')}
 G^{(+)}[{\bf x},\epsilon_{\rm C}({\bf x},t)] dt.
\end{eqnarray}
The condition for this to be a genuine stochastic differential equation is that 
the matrix
\begin{eqnarray}\Label{Q36}
\left(\begin{array}{cc} G^{(+)}[{\bf x},\epsilon_{\rm C}({\bf x},t)]
 & {iu\over\hbar}\alpha^*({\bf x})^2 
\\
-{iu\over\hbar}\alpha({\bf x})^2 & 
 G^{(+)}[{\bf x},\epsilon_{\rm C}({\bf x},t)]
 \end{array}\right).
\end{eqnarray}
Thus, although the Q-function always exists and is positive, this does not 
necessarily mean the stochastic differential equation is valid---see 
\cite{Gardiner1989ax} Ch.6. 



\subsection{Approximations and simplifications}
\Label{sec.Wigner.approx}
\subsubsection{Effect of the cutoff}
These equations are expressed in a simplified form, since by definition the 
condensate band contains a finite range of energies, and hence of wavelengths. 
 
This means that the Laplacian, the delta functions, and the form of the 
trilinear term are modified by a projection into this band, as explained in 
QKIII and QKV.  As well, the delta functions become delocalized, and the 
noise correlation functions are then well defined at $ {\bf x}= {\bf x}'$.  The 
actual implementation of this procedure is somewhat technical, but is 
essentially straightforward.

 A reasonable way to do this 
is to express the $ \delta({\bf x}-{\bf y})$ terms as a summation over 
a set of orthogonal condensate eigenfunctions.  This amounts to 
truncating an exact expression for a delta function at the cutoff energy 
$ E_{R}$.  But for sufficiently high energies the trap eigenfunctions 
approach the harmonic oscillator eigenfunctions, so we may instead use the 
expression in terms of harmonic oscillator eigenfunctions
$ y^*_{\bf n}({\bf x})$
\begin{eqnarray}\Label{C1}
\delta({\bf x}-{\bf y}) &\to &  \delta_{R}({\bf x},{\bf y})
\\ \Label{C2}
&=& \sum_{E({\bf n})< E_{R}}y^*_{\bf n}({\bf x})y_{\bf n}({\bf y})
\end{eqnarray}
This representation necessarily gives an added noise which is only significant 
for $ {\bf x},{\bf y}$ inside the region classically allowed  for energy equal 
to $ E_{R}$, and thus permits the use of fast Fourier transform methods.

\subsubsection{Effect of the cutoff on means and variances}
In application of the Wigner function to a field theory, the fact that each 
mode contributes an extra half of a quantum to averages---as shown in 
(\ref{28c})---means that there will be a contribution proportional to the 
number 
of modes retained below the cutoff.  For the Q-function the effect is similar, 
but the contribution is a whole quantum. 

Suppose   $ M$ modes contribute to the summation in (\ref{C2}).  If we define 
the particle number operators for the field theory and the various phase space 
representations
\begin{eqnarray}
\Label{C3}
\hat N &=& \int d^3{\bf x}\,\phi^\dagger({\bf x})\phi({\bf x})
\\
\Label{C4}
 N_P &=& \int d^3{\bf x}\,\alpha_P^*({\bf x})\alpha_P({\bf x})
\\
\Label{C5}
 N_W &=& \int d^3{\bf x}\,\alpha_W^*({\bf x})\alpha_W({\bf x})
\\
\Label{C6}
 N_Q &=& \int d^3{\bf x}\,\alpha_Q^*({\bf x})\alpha_Q({\bf x})
\end{eqnarray}
then the means and variances are related by
\begin{eqnarray}\Label{C7}
\langle N_P\rangle&=&\langle \hat N\rangle 
\\ \Label{C8}
\langle N_W\rangle &=& \langle \hat N\rangle +{M/2} 
\\ \Label{C9}
\langle N_P\rangle &=&\langle \hat N\rangle +M
\\
\Label{C10}
{\rm var}[ N_P] &=& {\rm var}[\hat N] - \langle \hat N\rangle 
\\ \Label{C11}
{\rm var}[N_W] & =& {\rm var}[\hat N] +{M/4}
\\ \Label{C12}
{\rm var}[N_Q] & =& {\rm var}[\hat N] +\langle \hat N\rangle +M
\end{eqnarray}
These results directly present the dilemma one is faced with in making a choice 
of representation.  
If we wish to consider only positive phase-space distributions, which therefore 
have a probabilistic interpretation, we can see immediately from these results 
that a P-function is only possible if there is at least a Poissonian number 
distribution---that is, the P-function does not always exist. (However the 
Positive P-representation can evade this restriction, at the cost of doubling 
the number of independent variables.)  In contrast, there is no restriction on 
the Wigner and Q-functions, which are known to exist for all physical states.

On the other hand, only the P-function has no ``vacuum noise'' contribution, 
proportional to $ M $, the number of modes, unlike the Wigner and Q-functions.  
For these, the existence of the ``vacuum noise'' contribution means that a 
simple interpretation of the stochastic differential equation (\ref{29}) as the 
equation of motion for the condensate wavefunction must require that the actual 
mean number of real particles $ \langle\hat N\rangle$ not be swamped by the 
vacuum contribution; that is we must require $  \langle\hat N\rangle\gg M $ at 
the very least.  This means that the mean occupation per mode is very much 
greater than 1, and that a \BEC\ must already be present.
For the P-function this is not a problem---$ M$ can be as large as we 
wish---even infinite if we desire. 

Finally, one must emphasize that the stochastic differential equation 
(\ref{29}) is approximate for the Wigner function interpretation, even with the 
appropriate noise properties, because of the neglect of third-order noise 
terms, 
as previously noted.  Fortunately, the condition that we van neglect these 
terms is the same as the condition that the vacuum modes not be dominant---that
the occupation per mode is high for the modes of interest. On the other hand
provided we have an initial distribution which is broader than the Poisson, 
and 
the temperature is sufficiently high for for the positivity condition of the 
noise matrix (\ref{P36}) to be satisfied, the P-function interpretation of the 
stochastic differential equations (\ref{29}) is valid and {\em exact}.

\subsection{The low temperature case}
For sufficiently low temperatures it is clear that all the transition rates
(\ref{23a}--\ref{23e}) are negligible, but the noise matrices
(\ref{P36},\ref{Q36})  no longer have positive eigenvalues.  This leaves us 
with only two feasible choices---either a Positive P-function interpretation,
which is numerically very difficult, or a Wigner function interpretation, which 
is approximate, but has the attractive property that {\em the noise 
contributions all vanish}, as can be seen from 
(\ref{30}--\ref{W34}).  This choice has been extensively investigated by 
Sinatra {\em et al.} \cite{Sinatra2001ax}.

\begin{figure}
\hskip 25mm\epsfig{file=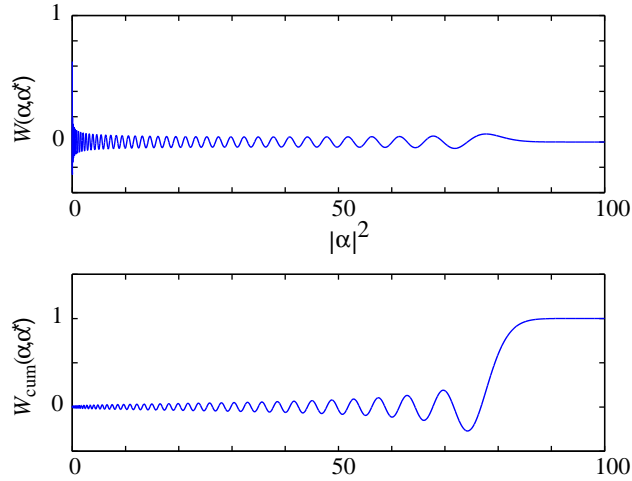}
\caption{a) The Wigner function for an $ N=80$ number state, 
and b) the corresponding cumulative Wigner function.\label{wigfig.eps} }
\end{figure}

\subsubsection{The stochastic Gross-Pitaevskii equation for a pure condensate}
To illustrate the procedure, we omit the noise and damping terms in (\ref{29}), 
and consider the resulting time dependent Gross-Pitaevskii for a random 
function $ \alpha({\bf x},t)$.  This takes the form

\begin{itemize}
\item[i)] $ \alpha({\bf x},t)$ satisfies the 
 Gross-Pitaevskii equation:
\begin{eqnarray}\Label{SGP0}
i\hbar{\partial\alpha\over \partial t } &=&
-{\hbar^2\over2m}\nabla^2\alpha+V_T({\bf x})\alpha
+u|\alpha |^2\alpha
.
\end{eqnarray}
The prescription of statistics in terms of
symmetrically ordered products as in (\ref{28a}--\ref{28c}) means that the 
initial conditions must be random even in the case of an initial pure state.
There are two possibilities, associated with the fact that the total number of 
particles is now exactly conserved, because of the vanishing of the 
terms (\ref{23a}--\ref{23e}). A straightforward interpretation would be to 
assume that the initial wavefunction of the condensate mode is 
$ \xi({\bf x},0)$
and use random initial 
conditions corresponding to all $ N$ particles being in the condensate, and 
none in the other $ M-1$ modes.  Unfortunately, the Wigner function for an $ N$ 
particle state involves the Laguerre polynomial:
\begin{eqnarray}\Label{IC1}
W_N(\alpha,\alpha^*) &=&
{2(-)^N\over\pi}\exp(-2|\alpha |^2){\rm L}_N(4|\alpha |^2),
\end{eqnarray}
and for $ N\ne 0$ this is a highly oscillatory function, as shown in 
Fig.~\ref{wigfig.eps}, 
which has no probability interpretation.
However, the cumulative integral
\begin{eqnarray}\Label{IC2}
W^{\rm cum}_N(\alpha,\alpha^*) &=&
\int d^2\alpha'\,\theta(|\alpha'|-|\alpha |)W_N(\alpha,\alpha^*) 
\end{eqnarray}
behaves very like a step-function, and this means that the mean values of any 
smooth functions of $ \alpha$, such as the first few powers, will be very well 
approximated by using a Gaussian approximation of the form
\begin{eqnarray}\Label{IC3}
\alpha &\to  &  \sqrt{n}\,e^{i\theta}
\end{eqnarray}
where $ \theta$ has a uniform distribution on $ (0,2\pi)$ and $ n $ is a real 
Gaussian random variable with
\begin{eqnarray}\Label{IC4}
\langle n\rangle &=& N+1/2
\\ \Label{IC5}
{\rm var}[n] &=& 1/4.
\end{eqnarray}
It should be borne in mind that in practice we are quite unable to even 
contemplate measuring any moments of $ \alpha({\bf x})$ higher than the fourth, 
and this form does correctly give all moments up to and including the fourth 
order.
\item[ii)]  The remaining modes would contain no particles, and thus for these 
the Wigner function (\ref{IC1}) is to be evaluated with $ N=0$, which 
is then is positive and Gaussian.

\item[iii)]
This means that we set
\begin{eqnarray}\Label{SGP1}
\alpha({\bf x},0) &=& \sqrt{n }e^{i\theta}\xi({\bf x},0) + 
\sum_k\xi_k({\bf x},0)\alpha_k
\\
 &=& \sqrt{n }e^{i\theta}\xi({\bf x},0) + \varphi({\bf x},0).
\end{eqnarray}
Here $ \alpha_k$ are complex Gaussian random variables, independent of each 
other, such that
\begin{eqnarray}\Label{IC6}
\langle\alpha_k\rangle &= &\langle\alpha_k^*\rangle=0
\\
\langle\alpha_k^2\rangle &= &\langle{\alpha_k^*}^2\rangle=0
\\
\langle\alpha_k^*\alpha_k\rangle&=&1/2.
\end{eqnarray}
It follows that that $ \varphi({\bf x},0)$ is a Gaussian random function with 
statistics given by 
\begin{eqnarray}\Label{WIV}
\langle \varphi({\bf x},0)\rangle &=& 0
\\ \Label{WV}
\langle \varphi({\bf x},0)\varphi({\bf y},0)\rangle &=& 0
\\ \Label{WVI}
\langle \varphi^*({\bf x},0)\varphi^*({\bf y},0)\rangle &=& 0
\\ \Label{WVII}
\langle \varphi^*({\bf x},0)\varphi({\bf y},0)\rangle &=&
{1\over 2}\left[\delta_{R}({\bf x},{\bf y}) - \xi^*({\bf x},0)\xi({\bf y},0)
\right]
\end{eqnarray} 

\item[iv)] The statistics of
the initial random condensate wavefunction
are Gaussian, with the non-zero means and correlations given by
\begin{eqnarray}\Label{SGP2}
\langle\alpha({\bf x},0)\rangle &=& \xi({\bf x},0)
\\ \Label{SGP3}
\langle\alpha^*({\bf x},0)\alpha({\bf y},0)\rangle &=&
N\xi^*({\bf x},0)\xi({\bf y},0)
 +{1\over 2}\delta_R({\bf x},{\bf y})
\end{eqnarray}
\item[v)] If there are $ M$ wavefunctions $ y_n$ included in the summations
(\ref{C2}), and hence $ M-1$ terms in the summation over $ k$ in
(\ref{SGP1}), then the mean number of particles $ N_W$ corresponding 
to the ensemble in (\ref{SGP2},\ref{SGP3}) is given by
\begin{eqnarray}\Label{SGP4}
N_W &=& \int d^3{\bf x}\,\langle |\alpha({\bf x},0)|^2 \rangle
=
N+{M\over 2}
\end{eqnarray}
This can be seen correspond to $ N$ real particles, plus the effect of ``half a 
quantum'' per mode for the $ M$ modes included.

\item[vi)]
If one runs a simulation for a time, and wishes to extract condensate and 
non-condensate number from the resulting ensemble, then the condensate 
wavefunction is 
\begin{eqnarray}\Label{SGP5}
\Psi({\bf x},t) &=& \langle\alpha({\bf x},t)\rangle
\end{eqnarray}
and the condensate number is
\begin{eqnarray}\Label{SGP6}
N_{\rm C} &=& \int d^3{\bf x}\,|\Psi({\bf x},t) |^2.
\end{eqnarray}
The number of particles not in the condensate is
\begin{eqnarray}\Label{SGP7}
N_{\rm T} &=& \int d^3{\bf x}\,\langle |\alpha({\bf x},t) |^2\rangle
-N_{\rm C} -{M\over 2},
\end{eqnarray}
that is, the ``vacuum'' particles must be subtracted.
\end{itemize}
It is clear that the ``vacuum particles'' will contribute to the evolution by 
means of the nonlinear mixing arising from the particle interactions, and that 
their effect is quite possibly cutoff dependent.  This cutoff dependence must 
disappear when one includes the full coupling to the noncondensate band by 
restoring  the damping and noise terms in the second and third lines of 
(\ref{29}).

\subsubsection{Treatment of a non-pure condensate}
It is not difficult in principle to include a condensate and its 
quasiparticles, and this was done in the original treatment \cite{Steel1998b}.  
One simply expresses the operators $ \alpha_k$ in terms of the quasiparticle
operators $ \beta_k$ for the system in the Bogoliubov theory, and then then one 
assigns to these the appropriate mean values instead of (\ref{IC6})
\begin{eqnarray}\Label{quas1}
\langle\beta_k\rangle &= &\langle\beta_k^*\rangle=0
\\
\langle\beta_k^2\rangle &= &\langle{\beta_k^*}^2\rangle=0
\\
\langle\beta_k^*\beta_k\rangle&=&\bar n_k+1/2.
\end{eqnarray}
Of course there will be considerable effort involved in solving the appropriate 
Bogoliubov-de Gennes equations in order to even construct the operators and 
their corresponding wavefunctions.  However, this can be made much easier by 
using the method of  Sinatra \etal \cite{Sinatra2000ax}, as 
explained in their later paper \cite{Sinatra2001ax}.

\subsubsection{The choice of $ E_R$}
It is tempting to choose $ E_R$ to be so large that the 
non-condensate-band phase space density $ F({\bf K},{\bf x})$ is 
negligible, and all of the noise and damping terms in (\ref{29}) can 
be neglected.  However, this may not be possible, since this would 
mean that the higher energy parts of the non-condensate band would 
also need to have negligible occupation, and in this case the neglect 
of third order derivative terms would certainly not be permissible.  
Thus it turns out that the principal criterion for the choice of $ 
E_R$ should be that there is significant occupation of all quantum 
states up to this level.

\section{Approximate phenomenological equations}
\Label{secIV}
The methods outlined above are quite complicated, and do not entirely answer 
the need for a simple set of equations which can give a quick estimate of the 
effects of being studied.  Therefore we will consider in here a very 
simplified 
equation obtained by making some drastic approximations to the already 
approximate methodology we have developed.
\subsection{Mean value equations}
Let us therefore neglect the terms involving $ E^{\pm}$, for the reasons noted 
above in Sect.\ref{Conservation} and consider the 
resulting equations of motion for the mean values: 
\begin{eqnarray}\Label{24}
\bar\phi({\bf x},t)
& \equiv& {\rm Tr}\left\{\phi({\bf x},t)\rho_{\rm C}(t) \right\},
\\ 
\bar n_{\rm C}({\bf x},t) &\equiv&
{\rm Tr}\left\{\phi^\dagger({\bf x},t)\phi({\bf x},t)\rho_{\rm C}(t) \right\},
\\
\bar {\bf j}_{\rm C}({\bf x},t) &\equiv&
{\rm Tr}\left\{ {i\hbar\over 2m}
\left[\phi^\dagger({\bf x},t)\nabla\phi({\bf x},t)
- \nabla[\phi^\dagger({\bf x},t)]\phi({\bf x},t)\right]
\rho_{\rm C}(t) \right\},
\nonumber \\
\end{eqnarray}
The resulting equation includes a modified \GPE, and a local growth equation.  
We include the relationship between the backward and forward rates $ G^{\pm}$, 
which arises from the definitions (\ref{23a},\ref{23b})
\begin{eqnarray}\fl\Label{25}
 G^{(-)}[{\bf x},\epsilon_{\rm C}({\bf x},t)] =
e^{ \epsilon_{\rm C}({\bf x},t)-\mu\over kT}
 G^{(+)}[{\bf x},\epsilon_{\rm C}({\bf x},t)]
\end{eqnarray}
\begin{eqnarray}\fl\Label{26}
{\partial\bar\phi({\bf x},t) \over\partial t } =
{i\over\hbar}  \left\{
{\hbar^2\over2m}\nabla^2\bar\phi( {\bf x},t)-V_T({\bf x})\bar\phi( {\bf x},t)
-2u \rho_{\rm NC}({\bf x},t)\bar\phi( {\bf x},t)
 +u\left\langle\phi^\dagger({\bf x})\phi({\bf x})^2\right\rangle 
\right\}
\nonumber\\ \fl \qquad\qquad
+ \left\{ G^{(+)}[{\bf x},\epsilon_{\rm C}({\bf x},t)]
\left(1 - e^{ \epsilon_{\rm C}({\bf x},t)-\mu\over kT}\right)
          -  M[{\bf x}]\right\} \bar\phi({\bf x},t),
\\ \fl \Label{27}
{\partial \bar n_{\rm C}({\bf x},t) \over\partial t} =
{\bf \nabla}\cdot\bar{\bf j}({\bf x},t)
+ 2 G^{(+)}[{\bf x},\epsilon_{\rm C}({\bf x},t)]\left\{
\big(1 - e^{ \epsilon_{\rm C}({\bf x},t)-\mu\over kT}\big)
\bar n_{\rm C}({\bf x},t) +1\right\} 
\end{eqnarray}
Notice that
\begin{itemize}
\item[i] 
The terms involving $ G^{\pm}$ give local growth or decay of 
$ \bar\phi$ and $ \bar n_{\rm C}$, and they wall also cause some dephasing, as 
was analysed in QKIII and QKIV.  When the forward and backward rates balance.
\item[ii] 
The spontaneous term---the $ +1$ in (\ref{27})---can initiate the 
condensate growth, while the term proportional to $ \bar n_{\rm C}$ gives the 
difference between stimulated growth and decay, which occur only with nonzero 
$ 
\bar n_{\rm C}$.
\item[iii]
The coefficient $ M[{\bf x}]$ has no effect on the density 
$ \bar n_{\rm C}$---it is a pure dephasing term of the kind well known in 
quantum optics and, as can bee seen from (\ref{25}), it makes the amplitude of 
the coherent component $ \bar\phi$ decay. Since it has no effect on the total 
number $ \bar n_{\rm C}$, this means that the the effect is to transfer 
particles from the coherent component into the incoherent or thermalized 
component of the condensate band.

\item[iv)]
This kind of equation cannot be expected to give a good description of the full 
condensate growth process, from no atoms in the condensate up to a fully 
developed condensate in equilibrium with a thermal vapour. As shown in 
\cite{Gardiner1998b,Lee2000ax,Davis2000ax,Bijlsma2000ax} the full growth theory 
requires a much more detailed description of the kinetics of the thermal cloud, 
and one cannot simply assume the thermal cloud is always in equilibrium at a 
definite temperature and chemical potential. However, in \cite{Gardiner1998b}
we noted that such a simple picture is reasonably valid once the condensate is 
reasonably large, say 50\% of its final value, and in \cite{Davis2000ax} it was 
noted that the growth process is indeed well described as being at  definite 
temperature, but that the chemical potential one should use is an ``effective 
chemical potential'', evaluated from the lower lying energy levels of the 
thermal vapour.  Thus our description is probably a reasonable description of 
the process of matter exchange between condensate and thermal cloud in a 
situation not unreasonably far from equilibrium. 
\end{itemize}

\subsection{The phenomenological growth equation}
The hydrodynamic approximation in practice should be defined in terms of the 
mean wavefunction $ \bar\phi( {\bf x},t)$, so that, to the extent that we can 
neglect derivatives of $ n_{\rm C}({\bf x},t)$, we can write
\begin{eqnarray}\Label{2701}
\epsilon_{\rm C}({\bf x},t)& =& 
-\hbar{\partial\Theta({\bf x},t) \over\partial t }
\\
&\approx& {i\hbar\over\bar\phi({\bf x},t) }
{\partial\bar\phi({\bf x},t) \over\partial t }.
\end{eqnarray}
We further make three further approximations:
\begin{itemize}
\item[i)] The argument of the exponent
in (\ref{25}) is sufficiently small to use $ e^x\approx 1+x$. This is in 
practice almost always true.
\item[ii)] We factorize all averages of products of $ \phi$ operators.
\item[iii)]
We neglect completely the term $M[{\bf x}] \bar\phi({\bf x},t)$ in 
(\ref{26}), since it is expected to be small, and its effect is merely 
to cause a small change in $\mu$.
\end{itemize}
Doing these, we can replace (\ref{26}) by
\begin{eqnarray}\fl\Label{2702}
{\partial\bar\phi({\bf x},t) \over\partial t } =
{i\over\hbar}  \left\{
{\hbar^2\over2m}\nabla^2\bar\phi( {\bf x},t)-V_T({\bf x})\bar\phi( {\bf x},t)
-2u \rho_{\rm NC}({\bf x},t)\bar\phi( {\bf x},t)
-u\left |\bar\phi({\bf x})\right |^2\bar\phi( {\bf x}) 
\right\}
\nonumber \\ \fl \qquad\qquad
W^{+}
\left\{\mu\bar\phi({\bf x},t) 
- i\hbar{\partial\bar\phi({\bf x},t) \over\partial t }\right\}.
\end{eqnarray}
where, for consistency with the notation of our previous papers, we 
have written
\begin{eqnarray}\Label{2703}
W^{+}&\equiv &{ G^{(+)}[{\bf x},\epsilon_{\rm C}({\bf x},t)]}.
\end{eqnarray}
The function $ G^+$ is slowly varying in space and time, so that $W^+$ 
can probably be approximated by a constant, since the most essential 
time- and space-dependence has been included in the factor 
$ 
\mu\bar\phi({\bf x},t) 
- i\hbar{\partial\bar\phi({\bf x},t) /\partial t }$.
The correct equilibrium results from the vanishing of the 
coefficent of $W^+$, which ensures that the time dependence of the 
wavefunction is $\exp(-i\mu t/\hbar)$---the remainng terms in the 
equation then reduce to the time independent \GPE\ with chemical 
potential $\mu$.
Thus we get a stationary condensate wavefunction with the same 
chemical potential as the noncondensate.

\section{Application to vortex array stabilisation}
\Label{vortex.lattice}
It is numerically known 
\cite{Caradoc-Davies1999a,Caradoc-Davies2000a,Feder1999a,Feder1999b} 
that merely stirring a condensate described by the \GPE\ produces 
vortices, but these do not stabilise into an regular array of 
vortices---as is experimentally observed 
\cite{Chevy2000a,Madison2000a,Raman2001ax}---merely as the result of 
the ongoing progress of the solution of the time-dependent \GPE. To 
create a vortex array numerically, one solves the time-dependent \GPE\ 
with an imaginary time and an appropriately time-dependent added 
chemical potential, while constantly renormalizing the wavefunction so 
as to maintain a constant number of particles.  This process finds a 
state with a energy minimum, but of course the method is purely an 
artefact, and does not represent the true underlying physics.

However the 
form of the phenomenological growth equation (\ref{2702}) already 
includes what amounts to an imaginary time term, as well as a real 
time term.  
We first add an angular momentum term to transform to 
the rotating frame, and then use a simple form for $ W^+$ as given in 
\cite{Gardiner1997b}
\begin{eqnarray}\Label{simp7}
W^+(N) &\approx& g{4m (akT)^2 \over \pi \hbar^3}.
\end{eqnarray}
where $a$ is the scattering length for the nonlinear interaction term 
(so that $u= 4\pi a\hbar^2/m$), $k$ is Boltzmann's constant, and $T$ 
the temperature of the noncondensate).  We have included the 
correction factor $ g\approx 3$ to give an approximate match with the 
more detailed treatment, as suggested in \cite{Gardiner1998b}.  The 
physics of this situation is the growth of a condensate in a frame 
rotating with angular velocity $ \Omega$ about the $ z$-axis from a 
vapour cloud which is itself stationary in the rotating frame; thus in 
the laboratory frame, this is vortex nucleation from a rotating vapour 
cloud, such as has been experimentally implemented by the JILA group 
\cite{Haljan2001ax} with no rotating trap potential.

Thus we find that we get
\begin{eqnarray}\fl\Label{G8}
\left(i-\gamma\right)\hbar{\partial\psi\over\partial t}
&=&-{\hbar^2\over 2m}\nabla^2\psi
+V_T({\bf x})\psi+  u \bigl |\psi\bigr |^2 \psi 
-\Omega L_z\psi + i\gamma\mu_{\rm NC}{\psi}.
\end{eqnarray}
Here 
\begin{eqnarray}\Label{G9}
\gamma\equiv {4mg a^2k T \over \pi \hbar^2}.
\end{eqnarray}
and using the values $ a\approx 10^{-8}{\rm m}$ and $ g=3$, we find the factor
$ \gamma\approx 0.01$.

As noted in the previous section, the stationary solution of this 
eqaution will come from the equality of the coefficients of $\gamma$ 
on both sides, ensuring that the time dependence of the wavefunction 
in the rotating frame is 
$\exp(-i\mu_{\rm NC}t/\hbar)$, leaving the wavefunction to satisfy the 
stationary \GPE\ (modified by the angular momentum term) with chemical potential 
$\mu_{\rm NC}$. This stationary solution will be a vortex 
lattice when it exists.

More generally, we can consider a trap rotating with angular velocity
$\Omega$ nucleating from a cloud rotating with angular velocity  
$\alpha$.  In this case we get, in the frame rotating with the trap,
\begin{eqnarray}\fl\Label{G801}
\left(i-\gamma\right)\hbar{\partial\psi\over\partial t}
&=&-{\hbar^2\over 2m}\nabla^2\psi
+V_T({\bf x})\psi+ u \bigl |\psi\bigr |^2 \psi
 -{\Omega}{L_z}\psi
+ i\gamma
\left\{ \mu_{\rm NC}+({\alpha}-{\Omega}){ L_z}\right\}
\psi 
.
\nonumber 
\\ \fl &&
\end{eqnarray}
The first equation (\ref{G8}) is similar to that presented in the paper of 
Tsubota {\em et al.} \cite{Tsubota2001ax}, but there are significant 
differences:
\begin{itemize}
\item[i)] 
There is no physical justification given in \cite{Tsubota2001ax}, and 
therefore their results, though qualitatively attractive, cannot be 
accepted as an explanation of the vortex lattice stabilization 
process.  In contrast, our reasoning shows their choice of $ 
\gamma=0.03$ is of the same order of magnitude as that expected from 
quantum kinetic theory.

\item [ii)] 
They have a {\em real} term $- \mu\psi$ on the RHS, 
whereas this equation has an {\em imaginary} term 
$ i\gamma\mu_{\rm NC}{\psi}$.

\item[iii)] 
Their $ \mu$ is adjusted with time to preserve the number of atoms in 
the condensate.  Our $ \mu_{\rm NC}$ is the physical chemical 
potential of the surrounding vapour, for which we have no particular 
model at the moment.  Even in the experiments 
\cite{Madison2000a,Raman2001ax} where an almost pure condensate is 
stirred by a rotating trap potential, it is known that as the stirring 
starts, there is considerable heating, and then at the end possibly 
20\% of the atoms form a thermalized rotating cloud, which appears to 
rotate at a lower speed than the trap.  Thus we can expect that 
(\ref{G801})  should give a semiquantitative description of the actual 
process of vortex formation and decay when appropriate (possibly 
time-dependent) values of $\mu_{\rm NC}$ and $\alpha$ are chosen. 
(However, we note that recent experimental work \cite{Abo-Shaeer2001a}
suggests that the stabilization process is only weakly dependent on 
temperature, which is surprising, since the dissipation in our equation 
is represented by $\gamma$, which is proportional to temperature.
A resolution of this anomaly must await our more detailed calculations 
which are presently in progress.)

The basic mechanism arises from the spatially dependent local chemical 
potential of a random lattice of vortices.  The equilibrium lattice is 
characterized by a uniform local chemical potential, which must 
balance that of any surrounding vapour.  The phenomenological growth 
equation includes this irreversible process.
\end{itemize}
We can expect a model with fixed $ \mu_{\rm NC}$ to explain the actual 
observed dissipation of the vibration of vortices which is observed, 
but the full process of formation and heating could well be a 
formidable task, involving both kinetic and GP methodologies.  
Preliminary calculations have shown that qualitatively, the results 
are very similar to those of Tsubota {\em et al.} 
\cite{Tsubota2001ax}---that is, the formation and stabilization of a 
vortex lattice are observed in very much the same way. The numerical 
techniques necessary for both equations are almost identical
\newcommand{\func}[1]{{\rm #1}}
\newcommand{\text}[1]{\mbox{#1}}
\renewcommand{\vec}[1]{{\bf #1}}

\section{Application to hydrogen condensate system}
\subsection{Nonlinear losses}\label{secV}

\begin{figure}
\hskip 25mm\epsfig{file=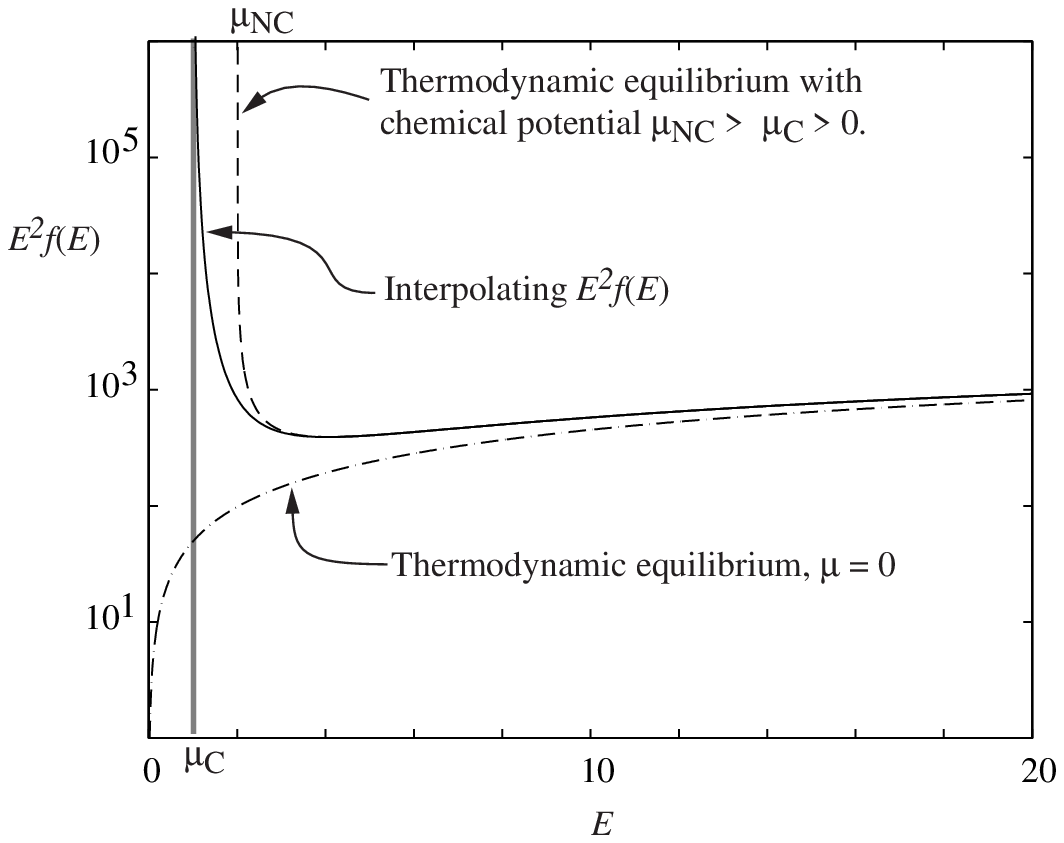,width=7cm}
\caption{\label{fig1} Illustration of the interpolation between a phase-space 
distribution 
form at high chemical potential $ \mu_{\rm NC}$ at high energies, and the 
phase-space distribution at low energies, whose chemical potential 
$ \mu_{\rm C}$is the same as that of the condensate} 
\end{figure}

The experiments on hydrogen 
\cite{Killian1998ax,Fried1998a,Killian2000a} give
rise to a system with a very large proportion of non-condensed hydrogen,
which feeds a relatively small condensate as the condensate atoms themselves
leave the condensate because of two-body dipolar relaxation. The data
indicate that the condensate appears to have a density profile which matches
that of a solution of the Gross-Pitaevskii equation. In contrast, the
non-condensed vapour has a profile which does not match that of a zero
chemical potential Bose-Einstein distribution, but instead seems to have
significantly more population at lower energies than was expected from the
zero chemical potential Bose-Einstein distribution. From the point of view
of the quantum kinetic theory of condensate growth processes the zero
chemical potential model chosen is not very appropriate, since any
condensate has a positive chemical potential, and that of the noncondensed
atoms must be even higher in order to maintain a net inflow of atoms into
the condensate to replace those which are continually leaving by two-body
processes. Thus, to the extent that a condensate in this situation can
indeed be described by a chemical potential $\mu_{\mathrm{C}}$, one expects
that the main body of the noncondensate would have a chemical potential $%
\mu_{\mathrm{NC}}>\mu_{\mathrm{C}}$, and that there would be a transition
region for the lower energy particles in the noncondensate. More precisely,
what one expects is a distribution function $f(E)$ of the form 
\begin{eqnarray}  \Label{GG1}
f(E) &=& {\frac{1}{\exp[(E - \mu(E))/kT] -1}}
\end{eqnarray}
where the \emph{energy-dependent chemical potential} $\mu(E)$ takes the
value $\mu_{\mathrm{NC}}$ for higher $E$, but approaches $\mu_{\mathrm{C}}$
as $E \to \mu_{\mathrm{C}}$---the situation is illustrated in figure \ref
{fig1}.

\subsubsection{Phenomenological growth equation including two-body losses}

We are most interested in the case of dynamic equilibrium, where the
stationary state of the condensate is maintained by the flow of the atoms
from the high chemical potential $\mu_{\mathrm{NC}}$ through the vapour at
chemical potential $\mu_{\mathrm{C}}$ and thence through the process of two
body dipolar relaxation to what amounts to chemical potential of $-\infty$,
since this process is irreversible. In this case we have to consider three
principal processes, namely

\begin{itemize}
\item[i)]  The \emph{linear growth} process, by which an atom leaves the
vapour and enters the condensate.

\item[ii)]  The \emph{linear loss process}, by which an atom returns to the
vapour from the condensate.

\item[iii)]  The \emph{nonlinear loss process}, by which pairs of atoms
leave the condensate by dipolar relaxation.
\end{itemize}

The first two of these terms are already included in the phenomenological
growth equation (\ref{2702}), The nonlinear loss is obviously modelled by a
term proportional to $-|\psi |^{2}\psi $, leading to a phenomenological
growth-loss equation for the hydrogen system: 
\begin{eqnarray}\fl\Label{AngA}
 i\hbar \dot{\psi}=-{\frac{\hbar ^{2}}{2m}}\nabla ^{2}\psi +V_{T}(\mathbf{%
x})\psi +u\bigl |\psi \bigr |^{2}\psi +i\hbar \left\{ W^{+}\left[ {\frac{\mu
_{\mathrm{NC}}\psi -i\hbar \dot{\psi}}{kT}}\right] -R|\psi |^{2}\psi
\right\} . 
\end{eqnarray}
It is a major advantage of the phenomenological approach which we have
developed that the nonlinear losses can be accommodated so
straightforwardly, with an effectively complex scattering length. 
The
simplicity of the modified Gross-Pitaevskii equation allows quantitative
investigation of questions which might well be difficult even to formulate,
if one had to proceed directly from first principles without the
phenomenological theory as an intermediate step. \ Yet among such questions
are those with potentially dramatic physical consequences, such as the one
we now pursue.

\subsection{Instability of the Thomas-Fermi stationary state under nonlinear
loss}

Under currently typical experimental conditions, the timescales of 
condensate growth and decay, including two-body loss, are much longer 
than those related to the harmonic trapping potential, the mean field 
energy $\mu_{\mathrm{C}}$, or the non-condensate chemical potential 
$\mu _{\mathrm{NC}}$.  \ To take advantage of this small ratio of 
frequency scales, it is usual to introduce length and time scales 
related to $\mu _{\mathrm{C}}$ (the healing length at peak density, 
and the mean field time scale); but since in the hydrogen context we 
want to see how the condensate maintains dynamic equilibrium in the 
presence of a much larger thermal cloud, we will consider $\mu 
_{\mathrm{NC}}$\ as the fixed parameter, which determines $\mu 
_{\mathrm{C}}$.  \ Since in dynamic equilibrium the two chemical 
potentials are comparable, this difference is not important.  So we 
will convert to dimensionless variables
\begin{eqnarray}
\tilde{t} &=&\frac{\mu _{\mathrm{NC}}}{\hbar }t \\
\mathbf{\tilde{x}} &=&\frac{\sqrt{m\mu _{\mathrm{NC}}}}{\hbar }\mathbf{x} \\
\tilde{V}(\mathbf{\tilde{x}}) &=&\mu _{\mathrm{NC}}^{-1}V(\mathbf{x}) \\
\tilde{u} &=&\mu _{\mathrm{NC}}^{-1}u\,.
\end{eqnarray}
This also motivates the dimensionless parameters 
\begin{eqnarray}
\varepsilon &=&2W^{+}\frac{\hbar }{kT} \\
\Lambda &=&\frac{kTR}{uW^{+}}
\end{eqnarray}
so that we can re-write (\ref{AngA}) as \ 
\begin{eqnarray}\Label{AngB}
i\dot{\psi}=-{\frac{1}{2}}\tilde{\nabla}^{2}\psi +\tilde{u}|\psi |^{2}\psi +%
\tilde{V}(\mathbf{\tilde{x}})\psi +{\frac{i\varepsilon }{2}}(\psi -i\dot{\psi%
}-\tilde{u}\Lambda |\psi |^{2}\psi )\;. 
\end{eqnarray}
Since $W^{+}$ is on the order of the Boltzmann scattering rate $\sigma \rho
_{tc}v_{T}$ for the thermal cloud, for the hydrogen condensate at
temperature approximately 50 $\mu $K this gives $\varepsilon \sim 10^{-6}$.
And with the published values for the two-body loss rates in this system,
the dimensionless parameter $\Lambda $ is on the order of $10$. \ 
The analysis below suggests that this is quite large enough to raise serious questions about the stability of the Thomas-Fermi stationary state; and it is possible that the large occupations in the lower quasiparticle levels may mean that $\Lambda\sim 10$ is an underestimate.

In the limit where $\varepsilon \rightarrow 0$, we obviously obtain 
the standard Gross-Pitaevskii equation.  \ Since the trapping 
potential varies very slowly on the healing length scale, in this 
limit the Thomas-Fermi approximation is excellent.  \ If $\varepsilon 
$ is of order unity or larger, the Thomas-Fermi stationary solution 
breaks down badly, and it may not even be possible to find a 
stationary solution to (\ref{AngB}).  But in the experimental regime 
of small $\varepsilon $, there are indeed stationary solutions in 
which the density profile is very close to the standard Thomas-Fermi 
form.  \ However, \textit{unless }$\Lambda $ is small enough\textit{, 
they are dynamically unstable.} By itself, the two-body loss term 
drives the total number of particles towards an equilibrium value.  
And it even tends to correct against collective excitations of the 
steady state condensate.  In bulk, it is easy to show that this 
corrective effect strongly stabilizes against perturbations; but as we 
show below, in a harmonic trap it overcorrects, and the collective 
excitations grow in amplitude.  Numerical integration reveals that 
these excitations continue in `boom-and-bust' cycles (where `bust' 
means a collapse of central density to a fraction of its maximum 
value); but in this regime, the assumptions that justify the 
phenomenological mean field theory may well be breaking down.  \ In an 
actual experiment, it is not clear whether we should expect coherent 
but non-stationary states like those assumed in our theory, or some 
kind of quasi-condensate with degraded phase coherence, or merely 
accelerated decay of the condensate.

\subsubsection{Stationary states}

We begin by extending the\ Thomas-Fermi stationary state to first order in $%
\varepsilon $. Since it is still true that $\tilde{V}$ is a slowly varying
function, we are in the hydrodynamic limit, and we can approximate 
(\ref{AngB}) by the
hydrodynamic equations for $\psi =\sqrt{\rho }e^{i\theta }$ and $\vec{v}=%
\vec{\nabla}\theta $, namely 
\begin{eqnarray}\Label{chd1}
\dot{\rho} &=&-\mathbf{\tilde{\nabla}}\cdot (\rho \mathbf{\tilde{\nabla}}%
\theta )+\varepsilon (1+\dot{\theta}-\tilde{u}\Lambda \rho )\rho  
 \\ \Label{chd2}
\dot{\theta} &=&-{\frac{1}{2}}|\mathbf{\tilde{\nabla}}\theta |^{2}-\tilde{u}%
\rho -\tilde{V}(\mathbf{\tilde{x}})\;,
\end{eqnarray}
where we drop a term $-\varepsilon \dot{\rho}/(4\rho )$ in the second
equation. \ (We drop this term because, even when below we allow
time-dependent density perturbations around the stationary state, this term
will be of order $\varepsilon ^{2}$. \ It does become large as we approach
the Thomas-Fermi surface, but this only leads to a small correction to the
usual boundary layer theory that must be applied to match the hydrodynamic
approximation to the near-surface region.) \ A stationary solution allows $%
\dot{\theta}=-\mu _{\mathrm{C}}/\mu _{\mathrm{NC}}$ as the only time
dependence in $\psi $, giving 
\begin{eqnarray}\Label{ssol1}
\vec{\nabla}\cdot (\rho \vec{v}) &=&\varepsilon (1-\mu _{\mathrm{C}}/\mu _{%
\mathrm{NC}}-\tilde{u}\Lambda \rho )\rho 
\\ \Label{ssol2}
\frac{\mu _{\mathrm{C}}}{\mu _{\mathrm{NC}}} &=&\tilde{u}\rho +\left( \frac{%
\hbar \omega _{r}}{\mu _{\mathrm{NC}}}\right) ^{2}\frac{\tilde{r}^{2}}{2}%
+\left( \frac{\hbar \omega _{z}}{\mu _{\mathrm{NC}}}\right) ^{2}\frac{\tilde{%
z}^{2}}{2}+\mathcal{O}(\varepsilon ^{2})
\end{eqnarray}
for cylindrical co-ordinates and an axisymmetric harmonic trap. \ So to
first order in $\varepsilon $, we have the familiar Thomas-Fermi density
profile for an axially symmetric harmonic trap: 
\begin{equation}
\rho ={\frac{1}{\tilde{u}}}\left[ \tilde{\mu}-\frac{\tilde{\omega}_{r}^{2}%
\tilde{r}^{2}}{2}-\frac{\tilde{\omega}_{z}^{2}\tilde{z}^{2}}{2}\right] \,;
\end{equation}
where we introduce the dimensionless trap frequencies 
$\tilde{\omega}_{j}\cg{\equiv \omega_j/\mu_{{\rm NC}}}$
and condensate chemical potential 
$\tilde{\mu}\cg{\equiv \mu/\mu_{{\rm NC}}}$. 
But there is now also a
velocity field $\mathbf{\tilde{\nabla}\theta }$\ of order $\varepsilon .$

Since a very small velocity field is difficult to observe, we are mainly
interested in the density profile. \ Actually finding $\vec{v}$ to first
order is therefore only indirectly necessary (it fixes $\mu _{\mathrm{C}}$);
but it is not so hard. Assume an ansatz of the form 
\begin{eqnarray}
\frac{\partial \theta }{\partial \tilde{r}} &=&\varepsilon \tilde{r}%
(a_{r}+b_{rr}\tilde{r}^{2}+b_{rz}\tilde{z}^{2}) \\
\frac{\partial \theta }{\partial \tilde{z}} &=&\varepsilon \tilde{z}%
(a_{z}+b_{zr}\tilde{r}^{2}+b_{zz}\tilde{z}^{2})
\end{eqnarray}
for unknown constants $a_{j},b_{jk}$. Since $\mathbf{\tilde{\nabla}}\theta $
must be irrotational, $b_{rz}=b_{zr}$. This leads to six equations for only
five free parameters in the ansatz, and so forces a specific value on the
heretofore undetermined term in $\rho $, namely the condensate chemical
potential $\tilde{\mu}$ (which makes sense, since we do expect the growth
and loss terms to determine $\mu _{\mathrm{C}}$). \ \ The result for the
axisymmetric trap is 
\begin{eqnarray}\Label{blah1}
\frac{\partial \theta }{\partial \tilde{r}} &=&-{\frac{\varepsilon }{7}}%
\frac{(7\tilde{\omega}_{r}^{2}+4\tilde{\omega}_{z}^{2})(\tilde{\mu}-\tilde{%
\omega}_{r}^{2}\tilde{r}^{2}/2)-\frac{11}{2}\tilde{\omega}_{z}^{2}\tilde{%
\omega}_{r}^{2}\tilde{z}^{2}}{5\tilde{\omega}_{z}^{2}+6\tilde{\omega}_{r}^{2}%
}\tilde{r}  \\
\Label{blah2}
\frac{\partial \theta }{\partial \tilde{z}} &=&-{\frac{\varepsilon }{7}}%
\frac{(4\tilde{\omega}_{r}^{2}+7\tilde{\omega}_{z}^{2})(\tilde{\mu}-\tilde{%
\omega}_{z}^{2}\tilde{z}^{2}/2)-\frac{11}{2}\tilde{\omega}_{z}^{2}\tilde{%
\omega}_{r}^{2}\tilde{r}^{2}}{5\tilde{\omega}_{z}^{2}+6\tilde{\omega}_{r}^{2}%
}\tilde{z}   \\
\Label{statres}                 
\tilde{\mu} &=&{\frac{1}{1+\frac{4}{7}\Lambda }}\;.  
\end{eqnarray}
The uniqueness of this ansatz solution can be proved straightforwardly in
the 1D limit, and also for spherical symmetry; we conjecture that it is more
generally unique. 
\begin{figure}[tbp]
\hskip 25mm\epsfig{file=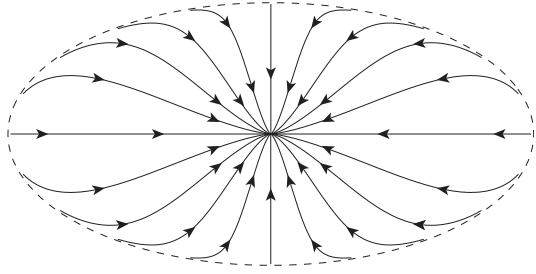,width=7cm}
\caption{\label{lowfiga.eps}Hydrodynamic flow lines in the presence of weak two-body losses,
for a harmonically trapped axisymmetric condensate with aspect ratio 2. \
The flow lines are as seen in a section parallel to the symmetry axis. \
Because of the symmetry of Eqn. (\ref{blah1},\ref{blah2}) under $r
\leftrightarrow z$,
this Figure is identical whether the condensate is oblate or prolate.}
\end{figure}

The results of (\cg{\ref{blah1}}--\ref{statres}) are somewhat complicated, but they make
excellent physical sense. \ Our expectation that the nonlinear losses will
lower $\mu _{\mathrm{C}}$ below $\mu _{\mathrm{NC}}$\ is borne out. \ And
the velocity field is inward from the edges of the condensate towards the
core. \ Since the loss rate scales faster with the density than the growth
rate, the periphery of the condensate receives from the thermal cloud more
particles than are needed to maintain the local condensate density, and so
it is able to donate a flux to the central regions, where loss rates exceed
growth. 
The velocity field becomes radial and proportional to $\rho $ in the limit
of spherical symmetry, but otherwise it has no simple description; see
\cg{Fig.\ref{lowfiga.eps} } for an illustrative example. 
In the limit $\omega _{z}\ll\omega_{r}$, we find $\partial _{z}\theta $ becoming independent of $r,$ and the
same statement with $r$ and $z$ interchanged is also true, so we do obtain
1D and 2D limits where we expect them. 

And it is always true that $\mathbf{%
\nabla \theta }\cdot \mathbf{\nabla }\rho =0$ at the edge of the
Thomas-Fermi cloud, which is required in the hydrodynamic approximation. \
(Density vanishing outside the TF surface means no flux can cross, since
gain and loss terms both vanish at zero density).

So with the addition of the slow flux from periphery to core, the
Thomas-Fermi solution is essentially maintained in the presence of slow
two-body losses. \ But is the Thomas-Fermi solution stable?

\subsubsection{Collective excitations}

To answer this question we need to find the frequencies of linear collective
excitations, by linearizing (\ref{chd1},\ref{chd2}) about the stationary state 
(\ref{ssol1},\ref{ssol2}). 
Using a method which can be dignified as multiple scale analysis,
but which is essentially the same as time-independent perturbation theory in
ordinary quantum mechanics, we compute the frequencies of the collective
modes to first order in $\varepsilon $. The first order corrections are in
general imaginary (as is not surprising given that our perturbations are
imaginary). What may be surprising is that if $\Lambda $ exceeds a threshold
of order unity, the amplitudes of some collective excitations \textit{grow},
indicating that the two-body losses cause an over-corrective instability.

The linearized hydrodynamic equations, to first order in $\varepsilon $, are 
\begin{eqnarray}
\delta\dot{\rho} &=& -\mathbf{\tilde{\nabla}}\cdot (\rho \,\mathbf{\tilde{%
\nabla}}\delta \theta +\delta \rho \,\mathbf{\tilde{\nabla}\theta }%
)+\varepsilon (1-\tilde{\mu}-2\Lambda \tilde{u}\rho )\delta \rho -\rho \delta\dot{\theta } 
 \nonumber \\
\delta\dot{\theta }  &=& -\tilde{u}\delta \rho -\mathbf{\tilde{\nabla}\theta 
}%
\cdot \mathbf{\tilde{\nabla}}\delta \theta \;.
\end{eqnarray}
We can combine these into a single second order equation for $\delta \theta $
(again dropping higher order terms in $\varepsilon $): 
\begin{eqnarray}\fl
\delta\ddot{\theta}=\tilde{u}\mathbf{\tilde{\nabla}}\cdot (\rho \mathbf{%
\tilde{\nabla}}\delta \theta )-\delta\dot{\theta}\,\mathbf{\tilde{\nabla}}%
^{2}\theta -2\mathbf{\tilde{\nabla}\theta }\cdot \mathbf{\tilde{\nabla}}%
\delta \dot{\theta}-\varepsilon \lbrack 1-\tilde{\mu}-(2\Lambda -1)\tilde{u}\rho
]\delta\dot{\theta}\;.
\end{eqnarray}
The idea now is to look for $\delta \theta =e^{i\Omega \tilde{t}}\delta
\theta (\mathbf{\tilde{x}})$, and expand $\Omega =\Omega _{0}+\varepsilon
\Omega _{1}+...$, and $\delta \theta =\delta \theta _{0}+\varepsilon \delta
\theta _{1}+...$. At zeroth order we have 
\begin{equation}\Label{zo}
-\Omega _{0}^{2}\delta \theta _{0}=\mathbf{\tilde{\nabla}}\cdot \left[
\left( \tilde{\mu}-\frac{\tilde{\omega}_{r}^{2}}{2}\tilde{r}^{2}-\frac{%
\tilde{\omega}_{z}^{2}}{2}\tilde{z}^{2}\right) \mathbf{\tilde{\nabla}}\delta
\theta _{0}\right] \;.  
\end{equation}
Then using the orthonormality of the eigenfunctions of the zeroth order RHS,
we obtain as in the time-independent perturbation theory of quantum
mechanics 
\begin{eqnarray}
-2\Omega _{0}\Omega _{1}\int_{TF}\!d^{3}\tilde{r}\,\delta \theta _{0}^{2}
&=&-i\Omega _{0}\int_{TF}\!d^{3}\tilde{r}[1-\tilde{\mu}-(2\Lambda -1)\tilde{u}\rho
]\delta \theta _{0}^{2}  
\nonumber\\ \Label{shift}
\Rightarrow \Omega _{1} &=&{\frac{i}{2}}{\frac{\int_{TF}\!d^{3}\tilde{r}\,[(1-\frac{%
10}{7}\Lambda )\tilde{\mu}-(1-2\Lambda )V(\vec{r})]\delta \theta _{0}^{2}}{%
\int_{TF}\!d^{3}\tilde{r}\,\delta \theta _{0}^{2}}}\;,  
\end{eqnarray}
where the integral is over the Thomas-Fermi cloud. (The terms on the RHS
with $\mathbf{\tilde{\nabla}}\theta$ in them add up to $\mathbf{\tilde{\nabla}}\cdot
(\delta \theta _{0}^{2}\mathbf{\tilde{\nabla}}\theta)$, which by Stokes'
Theorem is a surface integral that, as we mentioned above, vanishes, since
the normal to the surface of the TF cloud is $\propto \mathbf{\nabla}\rho $.)

If $\Lambda <0.7$, then our stationary solution is stable because the
integrand in (\ref{shift}) is positive throughout the volume of integration;
but if $\Lambda >0.7$, then the integrand is negative in a volume around
origin, and so the integral might perhaps be negative at least for some
modes.  To investigate further we must learn the $\delta \theta _{0}$ for the various collective modes.

Solving (\ref{zo}) for the hydrodynamic modes of a harmonic trap can be
reduced to diagonalizing finite matrices using a Frobenius series approach.
In an axially symmetric trap it is not hard to find modes up to third order
polynomials analytically and exactly, and with spherical symmetry we can
find all the modes. \ In an extremely prolate (cigar-shaped) trap, we can
also find all the modes to zeroth order in $(\omega _{z}/\omega _{r})^{2}$.
\ We will proceed here with the spherical calculation. \ The extremely
prolate case is analysed in the Appendix, and here we will only quote its
results.

\subsubsection{Spherical trap}

In the spherical case it is convenient to solve (\ref{zo}) in spherical
polar co-ordinates, with the angular dependence of $\delta \theta _{0}$
obviously being a spherical harmonic $Y_{lm}$. \ With $\omega _{z}=\omega
_{r}=\omega $, define the rescaled spherical radius 
\begin{eqnarray}
\mathsf{r=}\frac{\tilde{\omega}}{\sqrt{2\tilde{\mu}}}\sqrt{\tilde{r}^{2}+%
\tilde{z}^{2}} 
\end{eqnarray}
and write $\delta \theta _{0}=Y_{lm}(\vartheta ,\phi )f_{nl}(\mathsf{r})$ so
that (\ref{zo}) becomes 
\begin{equation} \Label{eigerber}
\frac{2\Omega _{0}^{2}}{\tilde{\omega}^{2}}f_{nl}=\left[ l(l+1)(1-\mathsf{r}%
^{2})+\mathsf{r}^{-2}\partial _{\mathsf{r}}(\mathsf{r}^{4}-\mathsf{r}%
^{2})\partial _{\mathsf{r}}\right] f_{nl}.  
\end{equation}
(The reason for the additional subscript $n$ on $f_{nl}$ will appear
shortly.) Taking a Frobenius ansatz $f_{nl}=\sum a_{knl}\mathsf{r}^{k+2}$
for the radial dependence then yields the recursion relation 
\begin{eqnarray} \Label{recursphere}
&&\left[ \left( k+s+2\right) (k+s+3)-l(l+1)\right] a_{k+2,nl}
\nonumber  \\
&&\qquad=\left[ (k+s)(k+s+3)-l(l+1)-2\Omega _{0}^{2}/\tilde{\omega}^{2}\right]
a_{knl}.  
\end{eqnarray}
Regularity at the origin then forces $s=l,$ and further requires that $%
a_{knl}$ vanish for all $k$ with parity opposite to $l.$ \ Convergence as we
approach the Thomas-Fermi surface $\mathsf{r=1}$ forces the sequence of
non-zero $a_{knl}$ to terminate at some $k=n-l\geq 0$. \ This introduces the
quantum number $n$, which must have the same parity as $l$, and be greater
than or equal to it. We recover the familiar result 
\begin{eqnarray}
2\Omega _{0}^{2}=\tilde{\omega}^{2}\left[ n(n+3)-l(l+1)\right] ; 
\end{eqnarray}
the point of repeating the calculation to this point has been to recall the
recursion relation (\ref{recursphere}), from which we can evaluate the RHS
of (\ref{shift}). \ 

In our rescaled spherical co-ordinates the equation for $\Omega _{1}$
becomes 
\begin{eqnarray}
\Omega _{1}^{\rm sphere}={\frac{i\tilde{\mu}}{2}}\left[ 1-\frac{10}{7}\Lambda
-(1-2\Lambda ){\frac{\int_{0}^{1}d\mathsf{r\,}\,\mathsf{r}^{4}\,f_{nl}^{2}}{%
\int_{0}^{1}d\mathsf{r}\,\,\mathsf{r}^{2}f_{nl}^{2}}}\right] . 
\end{eqnarray}
We can evaluate the ratio of moments in this expression by using the
orthogonality under the weight $\mathsf{r}^{2}$of the eigenfunctions of (\ref
{eigerber}) of different $n$, together with a fact we can extract from the
recursion relation (\ref{recursphere}). \ Since the $f_{nl}$ are polynomials
of rank $n$ and definite parity, it follows that 
\begin{eqnarray}
f_{n+2,l} &=&\sum_{k=0}^{n-l}a_{k,n+2,l}\,\mathsf{r}^{k+l} \\
&=&\frac{a_{n+2-l,n+2,l}}{a_{n-l,nl}}\mathsf{r}^{2}f_{nl} \\
&&+\frac{a_{n+2-l,n+2,l}}{a_{n-l,nl}}\left( \frac{a_{n-l,n+2,l}}{%
a_{n+2-l,n+2,l}}-\frac{a_{n-2-l,nl}}{a_{n-l,nl}}\right) f_{nl} \\
&&+\sum_{j=0}^{n-2}C_{j}f_{jl}\text{ }
\end{eqnarray}
for some constants $C_{j}.$ \ This then means that 
\begin{eqnarray}
\mathsf{r}^{2}f_{nl}=\left( \frac{a_{n-2-l,nl}}{a_{n-l,nl}}-\frac{%
a_{n-l,n+2,l}}{a_{n+2-l,n+2,l}}\right) f_{nl}+\sum_{j\neq n}C_{j}f_{jl}, 
\end{eqnarray}
so that by orthogonality we have 
\begin{eqnarray}
{\frac{\int_{0}^{1}d\mathsf{r\,}\,\mathsf{r}^{4}\,f_{nl}^{2}}{\int_{0}^{1}d%
\mathsf{r}\,\,\mathsf{r}^{2}f_{nl}^{2}}} &=&{}\frac{a_{n-2-l,nl}}{a_{n-l,nl}}%
-\frac{a_{n-l,n+2,l}}{a_{n+2-l,n+2,l}} \\
&=&\frac{1}{2}\left[ 1+{\frac{(l+{\frac{1}{2}})^{2}}{(n+{\frac{1}{2}})(n+{%
\frac{5}{2}})}}\right]
\end{eqnarray}
where the evaluation follows straightforwardly from (\ref{recursphere}). \ 

This immediately yields our desired result 
\begin{equation}\Label{omeg1i}
\Omega _{1}^{\rm sphere}={\frac{i\tilde{\mu}}{4}}\left[ 1-{\frac{6}{7}}\Lambda
+(2\Lambda -1){\frac{(l+{\frac{1}{2}})^{2}}{(n+{\frac{1}{2}})(n+{\frac{5}{2}}%
)}}\right] \;.  
\end{equation}
>From this we can see that the so-called `surface' modes, with $l=n$, are
always stable. \ (The case $n=l=0$ is not an excitation, but simply the
stationary state, so this case does not count.) \ Since if $\Lambda <1/2$
then $\func{Im}\Omega _{1}^{\rm sphere}$ will be positive because $1-6\Lambda
/7>1-2\Lambda $, we need only look for instabilities in cases where $%
2\Lambda -1>0.$ \ In these cases it is clear that instability can only occur
for $\Lambda >7/6$, in which case it occurs first for smaller $l$ and larger 
$n$. \ Equation (\ref{omeg1i}) indicates that for $\Lambda >7/6$ there will
always be instabilities at sufficiently high $n$, but the hydrodynamic
approximation on which the equation is based will break down for $n$ larger
than some limit that depends on $\mu _{\mathrm{C}}$ and $\omega $, and so in
general the actual instability threshold will be somewhat higher. \ But for $%
l=0$, all modes with $n\geq 2$ will be unstable if $\Lambda $ exceeds only $%
77/64$, and so the instability threshold is not actually very sensitive to
post-hydrodynamic corrections.

\subsubsection{Extremely prolate axisymmetric trap}

Instability is not a pathology of an exactly spherical harmonic trap: it
occurs also in a quasi-1D trap (for which all of our calculations can be
repeated quite simply), and in an extremely prolate but hydrodynamically
three-dimensional trap (which is not the same thing). \ For the extremely
prolate axisymmetric trap the result obtained in the Appendix is 
\begin{eqnarray}\fl
\Omega _{1}^{\rm x-pro}={\frac{i\tilde{\mu}}{2}}\left[ {\frac{4}{7}}\Lambda
+(1-2\Lambda ){\frac{p^{2}+p(2n+3)+(n+2)(2n+1)}{(2p+2n+1)(2p+2n+5)}}\left( 1-%
{\frac{m^{2}}{n(n+2)}}\right) \right] .
\end{eqnarray}
where $|m|,n,p$ are whole numbers. \ The azimuthal quantum number $m$ must
have the same parity as $n,$ and $|m|\leq n$. In this case, all of the axial
modes ($n=m=0$) are stable even if $\Lambda \rightarrow \infty $. \ And the
dipole modes ($n=m=0,p=1$ and $|m|=n=1,p=0$) are not only stable, but have $%
\Omega _{1}$ independent of $\Lambda $. \ (This is also true in the
spherical case, and indeed for all harmonic traps, because the dipole modes
merely translate the entire condensate, and so do not disturb the local
balance between growth, loss, and flux.) \ \ Raising $|m|$ and $p$ tends to
stabilize; the most unstable modes will have $p=m=0$. \ For these modes we
can recognize that a dynamical instability occurs for $(6n+8)\Lambda >(7n+14)
$, which will occur at high $n$ for $\Lambda >7/6$, just as in the spherical
case. \ The same caution applies, that for $n$ too large the hydrodynamic
approximation breaks down. But for $\Lambda >7/5$, all modes with $%
p=m=0,n\geq 2$ will be unstable, and so again post-hydrodynamic effects
cannot shift the instability threshold very much. In the hydrogen condensate
experiment at MIT, where the trap has an aspect ratio of 400, the extremely
prolate limit definitely obtains, and with $\Lambda\sim 10$ there are clearly 
a lot of unstable
modes. \ We conclude that instability of the Thomas-Fermi stationary state
is a general phenomenon if there are sufficiently strong two-body losses. \
The important point here is that `sufficiently strong' means only that the
two-body rate constant be slightly greater than the growth rate constant.

\subsection{Implications}

Since the MIT hydrogen condensate experiments are well above the instability threshold, 
our phenomenological growth equation seems to indicate that the quasi-steady state of these condensates cannot be the Thomas-Fermi density profile with perfect phase coherence, which all other condensates exhibit well.  The problem is not so much that the Thomas-Fermi state should be oscillating unstably, but that the system cannot be expected to settle down to an unstable state in the first place.

How seriously should we take this conclusion, which is after all derived
from approximate solutions to a phenomenological theory?  We would argue that, 
even if the theory might not be quantitatively precise, the overcorrective instability which it predicts under two-body losses is a simple physical phenomenon.  When loss and gain processes scale differently with density, and density is inhomogeneous, particle flow is needed to maintain a steady state.  In effect the condensate velocity field is a control system that tries to maintain a constant density profile.  But moving atoms have momentum, and so the problem of overcorrection can obviously arise in this control system; whether this leads to instability is a detailed matter of length and time scales.  It is possible that our approximate theory exaggerates the overcorrection problem, or that it may be counteracted by factors not included in the theory; but it is a genuinely physical possibility, and not a mere artifact of our phenomenological approach.  

Among factors neglected in this Section,
which might tend to suppress instabilities in real systems, is the thermal
component of the velocity field. \ In the hydrogen experiments the thermal
fluctuations in the condensate velocity field can be estimated to be much
larger than the systematic flow which compensates for the two-body losses,
and so these should really be taken into account, by solving the full,
stochastic version of our modified Gross-Pitaevskii equation.  

It is a great merit of the phenomenological growth equation that it allows explicit calculation of the instability question, which would be difficult even to formulate directly from first principles.  But given the experimental data showing a high-density component of ultracold hydrogen, whose density profile seems consistent with those of condensates in the alkali vapours, the phenomenological growth equation cannot be correct when it concludes that the Thomas-Fermi ground state is impossible in hydrogen.  The results of this Section do indicate, however, that the balance between thermal fluctuations and dissipation (loss) must be much less trivial in hydrogen than in the other ultracold Bose gases.  Thus in this Section we have provided evidence that further development of non-equilibrium theory for cold, dilute, trapped bosons will provide insights into qualitatively new regimes, and not just small corrections to familiar results.  At the same time the stochastic Gross-Pitaevskii approach has proven itself as a workable tool which can, without requiring unreasonable effort, yield definite answers to nontrivial physical questions.

\section{Conclusion}
\label{secVI}

In this paper we have given indications of how a stochastic 
Gross-Pitaevskii equation should be defined and used in practice.  
There are two major issues to be resolved in this formulation
\begin{itemize}
\item[a)] 
What is the appropriate definition of the ``condensate 
wavefunction'' $\psi({\bf x},t)$ for which the stochastic \GPE\ should 
provide an equation of motion?  Our conclusion is that a multimode 
Wigner function definition is the most appropriate, and although this 
cannot be done exactly, or even for all possible quantum states, the 
approximations and restrictions which are thus implicit in this 
definition appear to us to be unlikely to cause in serious problems 
when applied to practical situations.  The most irksome complication 
is the necessity to include ``vacuum noise'' in the initial 
conditions, basically in order to make sure the Heisenberg uncertainty 
principle is not violated, as detailed in 
Sect.\ref{sec.Wigner.approx}.  However, at all but the very lowest 
temperatures this is unlikely to be an issue---the influence of 
invalid initial conditions will very soon be eliminated by the noise 
induced by the thermalized atoms.

\item[b)] The formulation a quantum kinetic master equation using 
the local formulation of energy and momentum conservation introduced 
by Zaremba \etal \cite{Zaremba1999a} creates a very much simpler 
master equation than that of our earlier formulations, and its 
transformation into the stochastic \GPE\ is then relatively painless.
There are, however, many technical details involving the best way of  
quantitatively specifying the noise coefficients which are still in 
need of refinement.  We do not see these as being too important 
quantitatively in practical situations, but they will eventually need 
to be attended to more precisely than we have done here.

\end{itemize}
The major result is that, subject to a number of caveats, including 
the correct choice of initial conditions, we can modify the 
Gross-Pitaevskii equation to include damping as a result of exchange 
with non-condensed atoms, in order to obtain a semiphenomenological 
description of the interacting systems which includes all of the 
quantum effects in at least an approximate form.  

 The development of the {\em phenomenological growth equation}, 
 is similar to one presented by Williams and Griffin 
 \cite{Williams2001b}, is the most useful immediate application of 
 this work.  This equation can be seen as a very simplified 
 description of the interaction of a condensate and a thermal cloud, 
 which includes the major processes in a consistent but  
 only semiquantitative way.  The formulae for vortex lattice 
 nucleation and stabilization given in Sect.\ref{vortex.lattice} are 
 new, and in work which we shall publish elsewhere, we will show how 
 a rich variety of behaviours arise out of this simple and elegant 
 formulation.

We have also exploited this equation investigate the description of 
growth and loss in a hydrogen condensate, and have shown that there 
are hitherto unremarked hydrodynamic instabilities in the Thomas-Fermi 
stationary state of a cold condensate suffering two-body losses.  We 
shall publish more detailed numerical studies of this system elsewhere.

\ack  We wish to thank our colleagues Rob Ballagh, Andreas Penckwitt, 
Adam Norrie , Peter Zoller
and numerous others 
at \BEC\ workshops in Trieste, Benasque and Leiden
for valuable 
discussions and feedback.  This work was supported by the Marsden Fund 
of the Royal Society of New Zealand under contract PVT-902.

\section*{Appendix: two body instability in extremely prolate traps}

\subsubsection*{Zeroth order modes and overlap integrals}
The first step is to introduce the rescaled co-ordinates 
\begin{eqnarray}
\tilde{r} &=&{\frac{\sqrt{2\tilde{\mu}}}{\tilde{\omega}_{r}}}R  \nonumber \\
\tilde{z} &=&{\frac{\sqrt{2\tilde{\mu}}}{\tilde{\omega}_{z}}}Z\;,
\end{eqnarray}
and define $2\Omega _{0}^{2}\equiv \nu \tilde{\omega}_{r}^{2}$, $\tilde{%
\omega}_{z}^{2}\equiv \eta \tilde{\omega}_{r}^{2}$, $\delta \theta
_{0}\equiv f(R,Z)e^{im\phi }$, so that our zeroth-order equation becomes 
\begin{eqnarray}
-\nu f(R,Z) &=&(\partial _{R}+R^{-1})(1-R^{2}-Z^{2})\partial
_{R}f-m^{2}R^{-2}(1-R^{2}-Z^{2})f  \nonumber \\
&&\qquad +\eta \partial _{Z}(1-R^{2}-Z^{2})\partial _{Z}f
\end{eqnarray}
and our frequency correction becomes 
\begin{equation}
\Omega _{1}={\frac{i\tilde{\mu}}{2}}\left[ {\frac{4}{7}}\Lambda +(1-2\Lambda
){\frac{\int_{0}^{1}\!dZ\int_{0}^{\sqrt{1-Z^{2}}}\!RdR\,(1-R^{2}-Z^{2})f^{2}%
}{\int_{0}^{1}\!dZ\int_{0}^{\sqrt{1-Z^{2}}}\!RdR\,f^{2}}}\right] \;.
\end{equation}

With an aspect ratio of 400 in the hydrogen experiment, we can certainly
stop at zeroth order in $\eta $. To do this it is convenient to introduce
yet a further new co-ordinate: $R\equiv \sqrt{1-Z^{2}}X$. (This necessitates
some care with expressing partial derivatives and measures in the new
variables, but the results can be checked by comparison with exact solutions
to the equations in the original variables, and we have done this for the
ten simplest modes.) \ In these final variables $X,Z$\ we have 
\begin{eqnarray}\fl\Label{Xeq}
-\nu f(X,Z) &=&(\partial _{X}+X^{-1})(1-X^{2})\partial _{X}f+m^{2}(1-X^{-2})
\nonumber   \\ \fl
&&\qquad +\eta (\partial _{Z}+{\frac{XZ}{1-Z^{2}}}\partial
_{X})(1-Z^{2})(1-X^{2})(\partial _{Z}+{\frac{XZ}{1-Z^{2}}}\partial _{X})f 
\\ \fl
\Omega _{1} &=&{\frac{i\tilde{\mu}}{2}}\left[ {\frac{4}{7}}\Lambda
+(1-2\Lambda ){\frac{\int_{0}^{1}\!dZ(1-Z^{2})^{2}\int_{0}^{1}\!XdX%
\,(1-X^{2})f^{2}}{\int_{0}^{1}\!dZ(1-Z^{2})\int_{0}^{1}\!XdX\,f^{2}}}\right]
\;.
\end{eqnarray}
It is therefore clear that $f(X,Z)=G(Z)F(X)+\mathcal{O}(\eta )$, $\nu =\nu
_{0}+\eta \nu _{1}+\mathcal{O}(\eta ^{2})$, with 
\begin{eqnarray}\label{etasep}
-\nu _{0}F=(\partial _{X}+X^{-1})(1-X^{2})\partial _{X}F+m^{2}(1-X^{-2})F\;. 
\end{eqnarray}
Writing $F(X)=\sum a_{k}X^{k}$, we get the recursion relation 
\begin{equation}\Label{recurF}
\lbrack (k+2)^{2}-m^{2}]a_{k+2}=[-\nu _{0}+k(k+2)-m^{2}]a_{k}\;.
\end{equation}
Since the series can only converge if it terminates, we conclude that $\nu
_{0}=n(n+2)-m^{2}$ for some integer $n$. Since $F$ must not blow up at $X=0$%
, we must eliminate negative powers of $X$; this requires that $a_{k}=0$ for
all $k<|m|$. This implies that, for a non-vanishing solution, $|m|\leq n$;
and since the recursion relation goes in steps of two, this further implies
that all non-vanishing solutions have $m$ the same parity as $n$.

We can use (\ref{etasep}) to show that solutions $F_{n,m}$ with different $n$ 
are orthogonal in the interval $X\in [0,1]$ with the weight $X$.
Consequently, we can evaluate moments such as $\int\!dX\,X^3F_{nm}^2/\int%
\!dX\,XF_{nm}^2$ using the recursion relation (\ref{recurF}), by observing
that 
\begin{eqnarray}\fl
X^2 F_{n,m} &=& \sum_k a_{n;k-2} X^k  \nonumber 
\\ \fl
&=& {\frac{a_{n,n}}{a_{n+2,n+2}}}F_{n+2,m} + \Bigl({\frac{a_{n,n-2}}{a_{n,n}}%
} - {\frac{a_{n+2,n}}{a_{n+2,n+2}}}\Bigr) F_{n,m} + \sum_l^{n-2} c_lF_{l,m}
\end{eqnarray}
for some $c_l$. This implies that 
\begin{equation}  \Label{momF}
{\frac{\int_0^1\!dX\,X^3 F_{n,m}^2}{\int_0^1\!dX\,X F_{n,m}^2}} = {\frac{%
(n+2)^2-m^2}{4 (n+2)}} - {\frac{n^2-m^2}{4n}} = {\frac{1}{2}}\Bigl[1+ {\frac{%
m^2}{n(n+2)}}\Bigr]\;.
\end{equation}

We can now go back to (\ref{Xeq}) to first order in $\eta $, and by
integrating over $X$ with $XF_{nm}$, project out the equation for $G(Z)$.
(We have to integrate by parts several times, and apply (\ref{momF}) as
well.) In fact we find a simpler equation for $g(Z)\equiv
G(Z)(1-Z^{2})^{-n/2}$ (which is reassuring since after all $%
X^{n}=R^{n}(1-Z^{2})^{-n/2}$, and so $F_{nm}(X)(1-Z^{2})^{n/2}g(Z)$ will be
regular at $R=1,Z\rightarrow \pm 1$ if $g$ is regular at $Z=\pm 1$). The
result is 
\begin{equation}
-\xi g=(1-Z^{2})g^{\prime \prime }-2(n+2)Zg^{\prime }
\end{equation}
where a fairly complicated bunch of terms involving $m$, $n$ and $\nu _{1}$
has been absorbed into the eigenvalue $\xi $. Similar analysis to that above
shows $\xi =p(p+2n+3)$ for non-negative integer $p$, determining $\nu _{1}$
and hence $\Omega _{0}$ as a function of the mode indices $m,n,p$. We have
thus found the complete hydrodynamic spectrum of an extremely elongated
harmonic trap.

By a similar analysis as already performed for the $F_{nm}$ above, we can
also compute the second moment of $g(Z)$ (which are orthogonal over the
interval $Z\in \lbrack -1,1]$ with the weight $(1-Z^{2})^{n+1}$). So we also
finally have the correction to the frequency to first order in $\varepsilon$: 
\begin{eqnarray}\fl \Label{yay} 
\Omega _{1} &=&{\frac{i\tilde{\mu}}{2}}\left[ {\frac{4}{7}}\Lambda
+(1-2\Lambda ){\frac{\langle p,n|(1-Z^{2})|p,n\rangle _{Z}}{\langle
p,n|p,n\rangle _{Z}}}{\frac{\langle n,m|(1-X^{2})|n,m\rangle _{X}}{\langle
n,m|n,m\rangle _{X}}}\right]  
\\ \fl
&=&{\frac{i\tilde{\mu}}{2}}\left[ {\frac{4}{7}}\Lambda +(1-2\Lambda ){\frac{%
p^{2}+p(2n+3)+(n+2)(2n+1)}{(2p+2n+1)(2p+2n+5)}}\left( 1-{\frac{m^{2}}{n(n+2)}%
}\right) \right] \;.
\end{eqnarray}

\section*{References}
\bibliographystyle{prsty}
\bibliography{StochGPE}

\end{document}